\global\def\draftcontrol{0}

   \def\versionno{ dfp}

\catcode`\@=11

\expandafter\ifx\csname draftcontrol\endcsname\relax\global\def\draftcontrol{0}
\fi

{\count255=\time\divide\count255 by 60
\xdef\hourmin{\number\count255}
\multiply\count255 by-60\advance\count255 by\time
\xdef\hourmin{\hourmin:\ifnum\count255<10 0\fi\the\count255}}
\def\draftdate{\number\month/\number\day/\number\year\ \ \ \hourmin }

\newcommand\makepapertitle{\par
  \begingroup
    \renewcommand\thefootnote{\@fnsymbol\c@footnote}%
    \def\@makefnmark{\rlap{\@textsuperscript{\normalfont\@thefnmark}}}%
    \long\def\@makefntext##1{\parindent 1em\noindent
            \hb@xt@1.8em{%
                \hss\@textsuperscript{\normalfont\@thefnmark}}##1}%
     \newpage
     \global\@topnum\z@   
     \@makepapertitle
     \thispagestyle{empty}\@thanks
  \endgroup
  \setcounter{footnote}{0}%
  \global\let\thanks\relax
  \global\let\makepapertitle\relax
  \global\let\@makepapertitle\relax
  \global\let\@thanks\@empty
  \global\let\@author\@empty
  \global\let\@date\@empty
  \global\let\@title\@empty
  \global\let\title\relax
  \global\let\author\relax
  \global\let\date\relax
  \global\let\and\relax
  \def\version{\let\version\@version\@gobble}
}
\def\@makepapertitle{%
  \newpage
   \ifnum\draftcontrol=1 {}
   \version\versionno
   \vskip 3em%
   \else
   \hfill\hbox to 3cm {\parbox{4cm}{\@pubnum}\hss}%
   \vskip 3em%
   \fi
   \begin{center}%
   \let \footnote \thanks
     {\LARGE {\@title}}%
     \vskip 1.5em%
     {\normalsize
       \lineskip .5em%
       \begin{tabular}[t]{c}%
         \@author
       \end{tabular}\par}%
     \vskip 1.5em%
     {\@bstract}%
     \end{center}%
     \vskip 1.5em
     \@date%
   \par
}

\gdef\@pubnum{}
\def\pubnum#1{%
  \gdef\@pubnum{#1}}

\gdef\@bstract{}
\def\Abstract#1{%
  \gdef\@bstract{%
   \parbox{\textwidth-0pc}{%
   \centerline{\bf Abstract}\penalty1000%
\kern.2cm%
\noindent
\renewcommand\baselinestretch{1.0}%
{#1}}}
}

\def\ps@paper{\let\@mkboth\@gobbletwo%
     \ifnum\draftcontrol=1
    \def\@oddfoot{\hbox to \textwidth{\tiny \versionno \hfil\tiny\draftdate}%
    \hskip -\textwidth \hbox to \textwidth{\hfil\rm\thepage\hfil}}%
     \else\def\@oddfoot{\hbox to \textwidth{\hfil\rm\thepage\hfil}}
     \fi
     \let\@evenfoot\@oddfoot
}

\def\body{\clearpage
          \pagestyle{paper}
    }

\def\@version#1{\ifnum\draftcontrol=1
\typeout{}\typeout{#1}\typeout{}
\vskip3mm\centerline{\hbox{\fbox{\normalsize{\tt DRAFT -- #1 -- }
                   {\draftdate}}}}\vskip3mm
\fi}
\let\version\@version
\long\def\eqlabel#1{\ifnum\draftcontrol=1
                    \tag@false  
                    \tag*{(\theequation) \hbox to -0.2cm{\hspace{0cm}\small{#1}\hss}}
                    \refstepcounter{equation}
                    \edef\@currentlabel{\theequation}
                    \ltx@label{#1}          
                    \else
                    \label{#1}
                    \fi
                    }
\let\st@bibitem\@bibitem
\let\st@lbibitem\@lbibitem
\ifnum\draftcontrol=1
  \def\@bibitem#1{%
    \st@bibitem{#1}\a@@label{#1}\ignorespaces}
  \def\@lbibitem[#1]#2{%
    \st@lbibitem[#1]{#2}\a@@label{#2}\ignorespaces}
  \def\a@@label#1{%
    \gdef\a@lab{\smash{\normalfont\small#1}}
    \ifvmode
      \if@inlabel
        \global\setbox\@labels\hbox{%
          \llap{\a@lab\let\a@lab\relax
                \kern\@totalleftmargin\kern\marginparsep}%
          \box\@labels}%
      \fi
    \fi}
\fi

\documentclass[12pt,letterpaper]{article}

\usepackage{amsmath,amssymb,array,calc,epsfig,rotating,bm,xcolor}
\usepackage[sort]{cite}
\usepackage{graphicx,esint,float}
\usepackage{psfrag,verbatim}
\usepackage[makeroom]{cancel}
\usepackage{xcolor,url}
\usepackage{hyperref}

\ifnum\draftcontrol=1
\fi

\renewcommand\baselinestretch{1.25}
\renewcommand\section{\@startsection {section}{1}{\z@}%
                                   {-3.5ex \@plus -1ex \@minus -.2ex}%
                                   {2.3ex \@plus.2ex}%
                                   {\normalfont\large\bfseries}}
\renewcommand\subsection{\@startsection{subsection}{2}{\z@}%
                                   {-3.25ex\@plus -1ex \@minus -.2ex}%
                                   {1.5ex \@plus .2ex}%
                                   {\normalfont\normalsize\bfseries}}
\renewcommand\subsubsection{\@startsection{subsubsection}{3}{\z@}%
                                   {-3.25ex\@plus -1ex \@minus -.2ex}%
                                   {1.5ex \@plus .2ex}%
                                   {\normalfont\normalsize\it}}
\renewcommand\paragraph{\@startsection{paragraph}{4}{\z@}%
                                   {-3.25ex\@plus -1ex \@minus -.2ex}%
                                   {1.5ex \@plus .2ex}%
                                   {\normalfont\normalsize\bf}}

\numberwithin{equation}{section}

\def\revise#1       {\raisebox{-0em}{\rule{3pt}{1em}}%
                     \marginpar{\raisebox{.5em}{\vrule width3pt\
                     \vrule width0pt height 0pt depth0.5em
                     \hbox to 0cm{\hspace{0cm}{%
                     \parbox[t]{4em}{\raggedright\footnotesize{#1}}}\hss}}}}

\newcommand\nxt[1]  {\\\fnxt#1}
\newcommand{\ie}{{\it i.e.,}\ }
\newcommand{\eg}{{\it e.g.,}\ }

\def\cala         {{\cal A}}

\def\cale         {{\cal E}}
\def\calf         {{\cal F}}
\def\calg         {{\cal G}}

\def\calh         {{\cal H}}

\def\calk         {{\cal K}}
\def\call         {{\cal L}}
\def\calm         {{\cal M}}

\def\calo         {{\cal O}}

\def\cals         {{\cal S}}
\def\calt         {{\cal T}}

\def\zet          {{\mathbb Z}}
\def\hw          {{\hat{\omega}}}

\def\del          {\partial}

\def\Im           {{\rm Im\hskip0.1em}}

\def\sqr#1#2{{\vcenter{\vbox{\hrule height.#2pt
 \hbox{\vrule width.#2pt height#1pt \kern#1pt
 \vrule width.#2pt}\hrule height.#2pt}}}}

\def\dd{\delta}

\def\aa1{\phi}
\def\cc1{\psi}

\def\tp{\tilde{p}}

\def\f0{\text{\boldmath$\varphi$}}
\def\h2{\mathfrak{h}}
\def\dfps{${{\rm DFP}_s}\ $}
\def\dfpb{${{\rm DFP}_b}\ $}

\catcode`\@=12

\begin{document}


\title{\bf Dynamical fixed points in holography}

\date{November 6, 2021}

\author{
Alex Buchel\\[0.4cm]
\it $ $Department of Physics and Astronomy\\ 
\it University of Western Ontario\\
\it London, Ontario N6A 5B7, Canada\\
\it $ $Perimeter Institute for Theoretical Physics\\
\it Waterloo, Ontario N2J 2W9, Canada
}

\Abstract{Typically, an interactive system evolves towards thermal equilibrium,
with hydrodynamics representing a universal framework for its
late-time dynamics. Classification of the dynamical fixed points
(DFPs) of a {\it driven} Quantum Field Theory (with time dependent
coupling constants, masses, external background fields, etc.) is
unknown. We use holographic framework to analyze such fixed points in
one example of strongly coupled gauge theory, driven by homogeneous
and isotropic expansion of the background metric --- equivalently, a
late-time dynamics of the corresponding QFT in
Friedmann-Lemaitre-Robertson-Walker Universe. We identify DFPs that
are perturbatively stable, and those that are perturbatively unstable,
computing the spectrum of the quasinormal modes in the corresponding
holographic dual.  We further demonstrate that a stable DFP can be
unstable non-perturbatively, and explain the role of the entanglement
entropy density as a litmus test for a non-perturbative
stability. Finally, we demonstrated that a driven evolution might not
have a fixed point at all: the entanglement entropy density of a
system can grow without bounds.
}

\makepapertitle

\body

\version\versionno
\tableofcontents

\section{Introduction and summary}\label{intro}

Thermodynamic equilibrium is an internal state of a system without the net macroscopic
flow of matter or energy. 
This equilibrium state is characterized by few time-independent, constant over the material
sample, properties: the energy density $\cale_{eq}$, the pressure $P_{eq}$,
the entropy density $s_{eq}$, the temperature $T$, etc.  
Isolated interacting quantum systems
typically\footnote{Some of the counterexamples are the integrable systems,
Fermi-Pasta-Ulam-Tsingou  problem \cite{2008PhT....61a..55D}, and gravitational
collapse in $AdS$ \cite{Balasubramanian:2014cja}.} reach thermodynamic
equilibrium at late times of their dynamical evolution \cite{PhysRevA.43.2046,PhysRevE.50.888}:
\begin{equation}
\lim_{\tau\to \infty} T_{\mu\nu}(\tau,\bm{x})={\rm diag}\left(\cale_{eq},P_{eq},\cdots P_{eq}\right)\,,
\end{equation}
where $T_{\mu\nu}$ are the component of the stress-energy tensor of the system
at time $\tau$ and the spatial location $\bm{x}$. Moreover, the approach to thermal equilibrium,
whenever the space-time gradients of the local thermodynamic observables are small
compare to the energy scale set by the local temperature, is universally governed
by the hydrodynamics \cite{Landau1987Fluid}. For example, in the absence of conserved charges,
given a time-like unit vector $u^\mu=u^\mu(\tau,\bm{x})$, 
the Landau frame stress-energy tensor decomposes as
\begin{equation}
T^{\mu\nu}=\cale\ u^\mu u^\nu+P\ \Delta^{\mu\nu}+\calt^{\mu\nu}\,,\qquad \calt^{\mu\nu}=-\eta\ \sigma^{\mu\nu}-\zeta\ \Delta^{\mu\nu}\
\left(\nabla\cdot u\right) \,,
\eqlabel{hydro}
\end{equation}
where $g_{\mu\nu}$ is the background metric, $\Delta^{\mu\nu}\equiv g^{\mu\nu}+u^\mu u^\nu$,
$\sigma^{\mu\nu}$ is some symmetric transverse traceless tensor of the first derivatives of $u^\mu$,
and  $\cale=\cale(\tau,\bm{x})$ is the local energy density.  The shear $\eta$
and the bulk $\zeta$ viscosities are functions of the local energy density,
and can be computed from the equilibrium two-point correlation functions of the
stress-energy tensor.  The local pressure $P=P_{eq}(\cale)$ is determined by
the equilibrium equation of state, and the
local entropy density $s=s(\cale)$ and the temperature $T=T(\cale)$ follow from the basic laws
of the equilibrium thermodynamics:
\begin{equation}
\cale+P= s\ T\,,\qquad d\cale = T\ ds\,.
\eqlabel{locst}
\end{equation}
The second law of thermodynamics postulates that the divergence of the entropy current
$\cals^\mu$ is non-negative under physical processes
\begin{equation}
\nabla\cdot \cals\ \ge 0\,.
\eqlabel{ds}
\end{equation}
There is no first-principle definition of $\cals^\mu$ far from equilibrium; in the
hydrodynamic approximation, \ie to the first-order in the gradients of the local
fluid velocity $u^\mu$, \cite{Bhattacharya:2011tra,Kovtun:2019hdm}
\begin{equation}
\cals^\mu=s\ u^\mu-\frac 1T\ \calt^{\mu\nu}u_\nu+\calo(\del^2 u)\,.
\eqlabel{defs}
\end{equation}
Conservation of the stress-energy tensor \eqref{hydro} then implies
\begin{equation}
T\ \nabla\cdot \cals = \zeta\ \left(\nabla\cdot u\right)^2+\frac \eta2 \sigma_{\mu\nu}\sigma^{\mu\nu}
+\calo(\del^3 u)\,,
\end{equation}
which is manifestly non-negative, provided the viscosities are positive. 
At thermal equilibrium the divergence of the entropy current \eqref{ds} vanishes. 

We can now provide a formal definition of a dynamical fixed point (DFP):
\bigskip

\noindent\fbox{%
    \parbox{\textwidth}{%
{\color{red} A {\it Dynamical Fixed Point} is an internal state of a
quantum field theory with spatially homogeneous and time-independent
one-point correlation functions of its stress energy tensor $T^{\mu\nu}$,
and (possibly additional) set of gauge-invariant local operators $\{\calo_i\}$,
\centerline{\bf \underline{and}}
strictly positive divergence of the entropy current at late-times:
\[
\lim_{\tau\to \infty} \biggl(\nabla\cdot \cals\biggr)\ > 0
\]
}
}%
}
\bigskip

Note that apart from the requirement of the strictly non-zero entropy production
rate at late times, characteristics of a DFP coincide with that of the
thermodynamic equilibrium. The 
late-time entropy production can arise
when a QFT is {\it driven} externally by varying in time 
the coupling constants of the relevant operators, masses, background fields
(\eg the space-time metric).
In this paper we study DFPs of a QFT in a cosmological background,
the de Sitter background spacetime in particular.

DFP classification necessitates the assignment of the entropy current
$\cals^\mu$ to a system,
defined for its arbitrary far-from-equilibrium configurations.
From the QFT perspective, this is an unsolved problem --- the progress
can be made though for theories with  a dual holographic gravitational
descriptions \cite{Maldacena:1997re,Aharony:1999ti}. In variety of
holographic
models \cite{Buchel:2014gta,Buchel:2017pto,Buchel:2019pjb}
it was rigorously proven that the comoving gravitational entropy density of
the apparent horizon, associated with some natural observer
(the spatial slicing), can not decrease with time, 
\begin{equation}
\frac{ds_{comoving}^{AH}}{d\tau}\ge 0\,.
\eqlabel{scomdot2}
\end{equation}
It is natural to identify this gravitational entropy density $s_{comoving}^{AH}$
as the comoving entropy density of a boundary QFT, $s_{comoving}$:
\begin{equation}
s_{comoving}\ \equiv\ s_{comoving}^{AH}\,.
\eqlabel{ident}
\end{equation}
If $a(\tau)$ is a scale factor of the QFT background $d$-dimensional
Friedmann-Lemaitre-Robertson-Walker (FLRW) Universe
\begin{equation}
ds_{d}^2=-d\tau^2+a(\tau)^2\ d{\bm x}^2\,,
\eqlabel{defflrm}
\end{equation}
the comoving $s_{comoving}$ and the physical $s$ entropy densities are
related as
\begin{equation}
s_{comoving}(\tau)\ =\ a(\tau)^{d-1}\ s(\tau)\,.
\eqlabel{defcomoving2}
\end{equation}
Thus, is we define the entropy current as
\begin{equation}
\cals^\mu= s(\tau)\ u^{\mu}\,,\qquad u^{\mu}\equiv (1,0,\cdots,0)\,,
\eqlabel{defs2}
\end{equation}
we find that
\begin{equation}
\nabla\cdot \cals
= \frac{1}{a(\tau)^{d-1}}\ \frac{d}{d\tau}\left(a(\tau)^{d-1} s(\tau)\right)=
\frac{1}{a(\tau)^{d-1}}\ \frac{d}{d\tau} s_{comoving}(\tau)\ \ge\ 0\,,
\eqlabel{ds2}
\end{equation}
due to \eqref{scomdot2}.
In \cite{Buchel:2017qwd}, the late-time, $\tau\to \infty$, limit of the
physical entropy $s(\tau)$, provided this limit exists, was called the
{\it vacuum entanglement entropy} (VEE) density
\begin{equation}
\lim_{\tau\to\infty} s(\tau)=s_{ent}\,.
\eqlabel{defsent3}
\end{equation}
Specializing to de Sitter Universe, \ie $a(\tau)=e^{H\tau}$,
we determine the late time entropy production rate as
\begin{equation}
\lim_{\tau\to \infty} \biggl(\nabla\cdot \cals\biggr)= (d-1)\ H\ s_{ent}\,.
\eqlabel{rate3}
\end{equation}
Due to \eqref{scomdot2}, the VEE is nonnegative; if it is strictly
positive, as in \cite{Buchel:2017pto,Buchel:2017lhu,Buchel:2019pjb},
it characterizes a DFP. 

Note that an interacting conformal field theory driven by \eqref{defflrm} can not have
a DFP. Indeed, a CFT$ _{d}$ dynamics in \eqref{defflrm} by a conformal transformation
is equivalent to a dynamics in Minkowski space-time.
Furthermore, the entropy production rate associated with the dynamics of the apparent horizon
is invariant under the conformal transformations \cite{Buchel:2017pto}.
Since a closed interacting theory in
Minkowski space-time is expected to equilibrate, 
\begin{equation}
\lim_{\tau\to\infty} \nabla\cdot \cals =0\,.
\end{equation}

As we explicitly demonstrate in the paper:
\begin{itemize}
\item A holographic  non-conformal
QFT can have several DFPs, depending
on the ratio of its mass-scale $\Lambda$ and the Hubble constant $H$.
\item As in \cite{Buchel:2019pjb}, distinct DFPs are characterized
by a pattern of the spontaneous global symmetry breaking ---
importantly, they have different vacuum entanglement entropy densities.
\item We present examples of DFPs that are perturbatively stable,
and those that are unstable to fluctuations of energy density and/or the
global symmetry order
parameter\footnote{This might have implications to 
cosmological model building: instabilities in late-time de Sitter
cosmology of a strongly coupled gauge theory can lead to observable
imprints in the cosmic microwave background.}.
\item Following \eqref{rate3},
when several DFPs are present given $\frac{\Lambda}{H}$,
the one with the larger $s_{ent}$ would lead at late times
to a larger comoving entropy production rate.
In our model, DFPs with the larger VEE density have  spontaneous symmetry
breaking\footnote{That a global symmetry broken DFP always
has a larger $s_{ent}$ is not true in general, see \cite{Buchel:2019pjb}
for a counterexample.}. The symmetry preserving DFP is perturbatively
stable to fluctuations of the symmetry breaking order parameter --- thus, to  
reach "more entropic'' symmetry broken DFP, one needs a large
enough amplitude of the symmetry breaking fluctuations. In other words,
a perturbatively stable DFP can be non-perturbatively unstable.
\item We demonstrate that a DFP of a non-conformal QFT
in de Sitter need not exist,
at least within controllable holographic framework: we identify
$\frac{\Lambda}{H}$ parameter range where initially arbitrarily
small amplitude  symmetry breaking fluctuations evolve the
(initially smooth) dual gravitational geometry
to a naked singularity. The physical entropy
density $s$ growth without bounds, implying that the limit
\eqref{defsent3} does not exist. 
\end{itemize}

The rest of the paper is organized as follows.
In the next section we introduce our holographic model ---
it was been extensively studied in holography in the past
\cite{Buchel:2009ge,Buchel:2009mf,Bosch:2017ccw,Buchel:2017lhu,Buchel:2020thm}.
While the model is a 'toy ' model of the holographic correspondence,
it has top-down holographic cousins, \ie
\cite{Buchel:2017pto,Buchel:2019pjb},
that share some similarities with respect to DFPs.
The advantage of the model is that it allows for
a relatively simple numerical simulation, enabling the
study of DFPs as late-time attractors of its dynamical evolution.
Numerical codes used present an adaptation of the codes
deployed in \cite{Bosch:2017ccw}  and \cite{Buchel:2017lhu}.
To keep the discussion self-contained, we review the
adapted codes in appendix \ref{num}.
Results are presented in section \ref{results}, with minimum of
technical details: the holographic dictionary of the model
is covered in appendix \ref{eoms}; appendices \ref{z2u} and \ref{z2b}
discuss symmetric and symmetry-broken DFPs of the model, including
the details necessary to compute the spectra of linearized fluctuations;
appendix \ref{enttheorem} contains the proof of \eqref{scomdot2}
for our model\footnote{It would be nice to prove the relevant
apparent horizon theorem in full generality, rather than on
case-by-case basis as done here, and in earlier
work \cite{Buchel:2014gta,Buchel:2017pto,Buchel:2017lhu,Buchel:2019pjb}.},
as well as the explicit expression for VEE densities of the model DFPs.
Finally, conclusions and future directions are
covered in section \ref{conclusions}.

The work reported here is heavily numerical.
Convergence tests of the numerical simulations are discussed in
appendix \ref{convtests}. Numerical construction of DFPs and
the computations of the spectra of the linearized fluctuations   
are done using Wolfram Mathematica, while the evolution codes
are implemented in FORTRAN. The fact that
perturbatively stable DFPs are attractors of the late-time dynamics,
including the approach rate to a DFP (as set by the lowest quasinormal mode
(QNM)
frequency of the linearized fluctuations),
is a highly nontrivial cross check between the two numerical platforms.
Likewise, there is an excellent agreement between the growth
rate of unstable (small amplitude) fluctuations in a simulation,
with the corresponding frequency of the unstable QNM of the DFP. 
Another nontrivial consistency check on numerics
is the high accuracy agreement between the two
different expressions for the vacuum
entanglement entropy densities of various DFPs, see appendix \ref{eed}.

\section{Holographic model}\label{model}

We restrict our attention to a simple holographic toy model of a $d=2+1$-dimensional massive $QFT_3$ 
with the effective dual gravitational action\footnote{We set the radius $L$ of an asymptotic $AdS_4$ geometry 
to unity.}:
\begin{equation}
\begin{split}
S_4=&S_{CFT}+S_{r}+S_i=\frac{1}{2\kappa^2}\int_{\calm_4} dx^4\sqrt{-\gamma}\left[\call_{CFT}+\call_{r}+\call_i\right]\,,
\end{split}
\eqlabel{s4}
\end{equation}
\begin{equation}
\call_{CFT}=R+6\,,\qquad \call_r=-\frac 12 \left(\nabla\phi\right)^2+\phi^2\,,\qquad 
\call_i=-\frac 12 \left(\nabla\chi\right)^2-2\chi^2-g \phi^2 \chi^2 \, ;
\eqlabel{lc}
\end{equation}
where we split the action into (a holographic dual to)  a $CFT_3$ part $S_{CFT}$; its deformation by a relevant
operator $\calo_\phi$,
\begin{equation}
L^2 m^2_\phi=-2=\Delta_\phi(\Delta_\phi-d) \;,
\eqlabel{mphi}
\end{equation}
and a sector $S_i$ involving an irrelevant operator $\calo_\chi$,
\begin{equation}
L^2 m^2_\chi=+4=\Delta_\chi(\Delta_\chi-d)\;,
\eqlabel{mchi}
\end{equation}
along with its mixing with $\calo_\phi$ under the renormalization-group flow,
represented by a bulk interaction $-g\phi^2\chi^2$.
We take bulk quantization so that  the scaling dimension of $\calo_\phi$ is
$\Delta_\phi=2$; the scaling dimension of $\calo_\chi$ is $\Delta_\chi=4$ .
In order to have asymptotically $AdS_4$ solutions, 
we assume that only the normalizable mode of $\calo_\chi$ is nonzero near the boundary.
The four dimensional gravitational constant $\kappa$ is related to the ultraviolet (UV) conformal fixed point
$CFT_3$  central charge $c$  as 
\begin{equation}
c=\frac{192}{\kappa^2}\,.
\eqlabel{c}
\end{equation}

The gravitational action \eqref{s4} has $\zet_2^\phi\times \zet_2^\chi$ discrete symmetry that acts as a parity 
transformation on the scalar fields $\phi$ and $\chi$ correspondingly.
The discrete symmetry $\phi\leftrightarrow -\phi$
is explicitly broken by the relevant deformation of the CFT, 
\begin{equation}
\calh_{CFT}\ \to \calh_{CFT}+\Lambda\ \calo_\phi \;,
\eqlabel{defformation}
\end{equation}
with $\Lambda$ being the deformation mass scale, 
while the $\chi\leftrightarrow -\chi$ 
symmetry is broken spontaneously whenever $\calo_\chi\ne 0$.
We are  interested in the holographic dynamics of the boundary theory
\eqref{s4}, driven by the homogeneous and isotropic expansion of its background metric
\begin{equation}
ds_{\del\calm_4}^2=-d\tau^2 +e^{2 H \tau}\ \left(dx_1^2+dx_2^2\right)\,,
\eqlabel{bmetric}
\end{equation}
with a constant Hubble parameter $H$.

A generic state of the boundary field theory with a gravitational dual \eqref{s4}, homogeneous and isotropic in the spatial
boundary coordinates $\boldsymbol{x}=\{x_1,x_2\}$, leads to a bulk gravitational metric ansatz\footnote{We use the characteristic formulation of the holographic dynamics
\cite{Chesler:2013lia}.}
\begin{equation}
ds_4^2=2 d\tau\ (dr -A d\tau) +\Sigma^2\ d\boldsymbol{x}^2\,,
\eqlabel{EFmetric}
\end{equation}
with the warp factors $A,\Sigma$ as well as the bulk scalar $\phi$ and $\chi$
depending only on $\{t,r\}$. We collect equations of motion obtained from
\eqref{s4}, within the bulk ansatz \eqref{EFmetric}, in appendix \ref{eoms}.
Appendix  \ref{eoms} explains how one extracts the $QFT_3$ observables:
the energy density $\cale(t)$, the pressure $P(t)$, and the expectation
values of the operators $\calo_\phi(t)$ and $\calo_\chi(t)$ from the
holographic bulk dynamics of the given initial state.  

Another important observable is the non-equilibrium entropy density $s(\tau)$.
This physical entropy density
should not to be confused with the comoving entropy density, corresponding to the
boundary QFT background space-time expansion \eqref{bmetric}:
\begin{equation}
s_{comoving}(\tau)\ =\ e^{2H\tau}\ s(\tau)\,.
\eqlabel{defcomoving}
\end{equation}
A holographic characteristic formulation of the boundary dynamics
\eqref{EFmetric} implies a preferred spatial slicing (a preferred observer) ---
we associate the comoving non-equilibrium entropy density
with the comoving Bekenstein-Hawking entropy density of the corresponding
apparent horizon, see appendix \ref{enttheorem}. As defined,
the comoving entropy density has the following properties:
\nxt It can only grow with time (see \cite{Buchel:2014gta,Buchel:2017pto,Buchel:2019pjb}
and section \ref{theo1}),
\begin{equation}
\frac{ds_{comoving}}{d\tau}\ge 0\,.
\eqlabel{scomdot}
\end{equation}
\nxt It coincides with the thermal entropy density at equilibrium.
\nxt In the adiabatic approximation, \ie $H\ll \Lambda$, and for initial configurations
close to thermal equilibrium, the comoving entropy
density production rate \cite{Landau1987Fluid,Landau:1980mil}
is governed by the hydrodynamics \cite{Buchel:2016cbj}:
\begin{equation}
\frac{d}{d\tau}\ \ln\ s_{comoving} \approx
(\nabla\cdot u)^2\ \frac{\zeta}{\cale +P} = (2H)^2\ \frac{\zeta}{\cale +P} \,,
\eqlabel{growthrate}
\end{equation}
where $\zeta$ is the bulk viscosity of the boundary QFT, and we identified
the homogeneous and isotropic expansion of the background space-time  \eqref{bmetric}
with the locally static hydrodynamic expansion $u^i=(1,0,0)$ with a nonzero
expansion rate $\Theta\equiv \nabla_i u^i=2 H$.

An initial state of a QFT,
for a specified driving rate $\frac H\Lambda$,
will evolve according to the gravitational dynamics as explained in appendix
\ref{eoms}. The late-time $H\tau\to \infty$ state is a DFP
provided the limit \eqref{defsent3} exists.

\section{Results}\label{results}

We choose
\begin{equation}
g=-100\,.
\eqlabel{gchoice}
\end{equation}
All the results reported depend on the choice of this
parameter in the effective action \eqref{lc}.

\subsection{$\zet_2^\chi$-symmetric DFP --- \dfps}

\begin{figure}[h]
\begin{center}
\psfrag{p}[cc][][1][0]{$\frac{\Lambda}{H}$}
\psfrag{e}[bb][][1][0]{$\frac{\cale}{cH^3}$}
\psfrag{r}[tt][][1][0]{$\frac{P}{cH^3}$}
\includegraphics[width=3in]{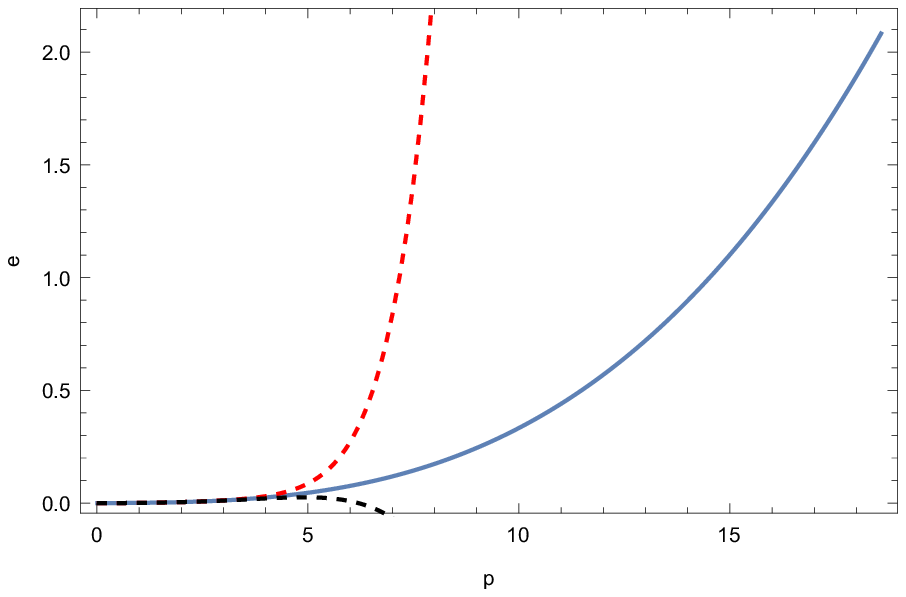}
\includegraphics[width=3in]{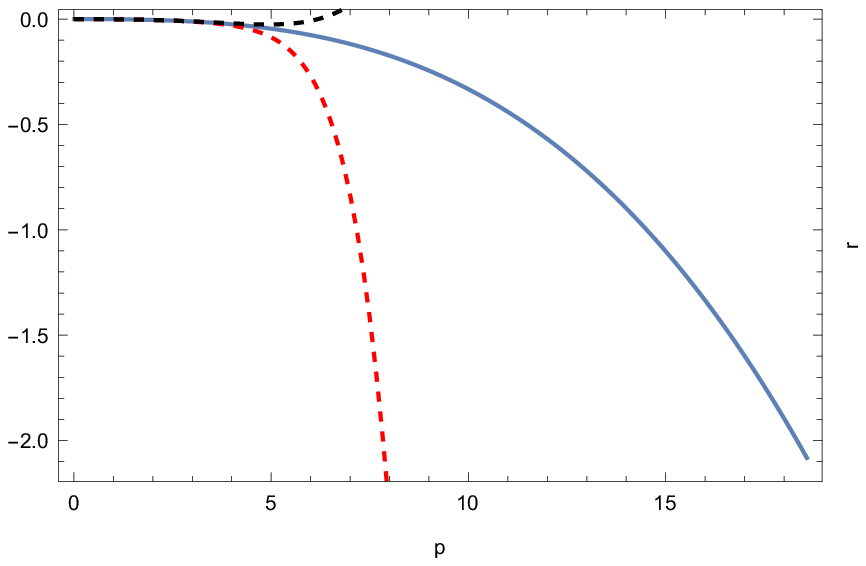}
\end{center}
  \caption{The energy density $\cale$
  (the left panel) and the pressure $P$ (the right panel)
  of the $\zet_2^\chi$-symmetric DFP (solid curves). The black/red
  dashed curves present the (successive) perturbative in the limit
  $\frac{\Lambda}{H}\to 0$ approximations to the observables.}
 \label{epunbroken}
\end{figure}

\begin{figure}[h]
\begin{center}
\psfrag{p}[cc][][1][0]{$\frac{\Lambda}{H}$}
\psfrag{o}[bb][][1][0]{$\frac{\calo_\phi}{cH^2}$}
\psfrag{e}[tt][][1][0]{$\frac{s_{ent}}{4\pi cH^2}$}
\includegraphics[width=3in]{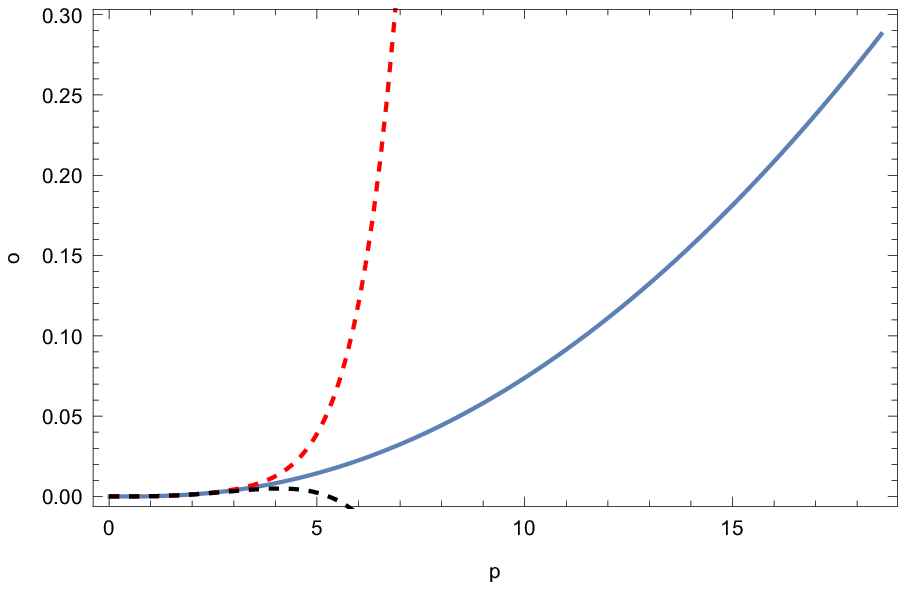}
\includegraphics[width=3in]{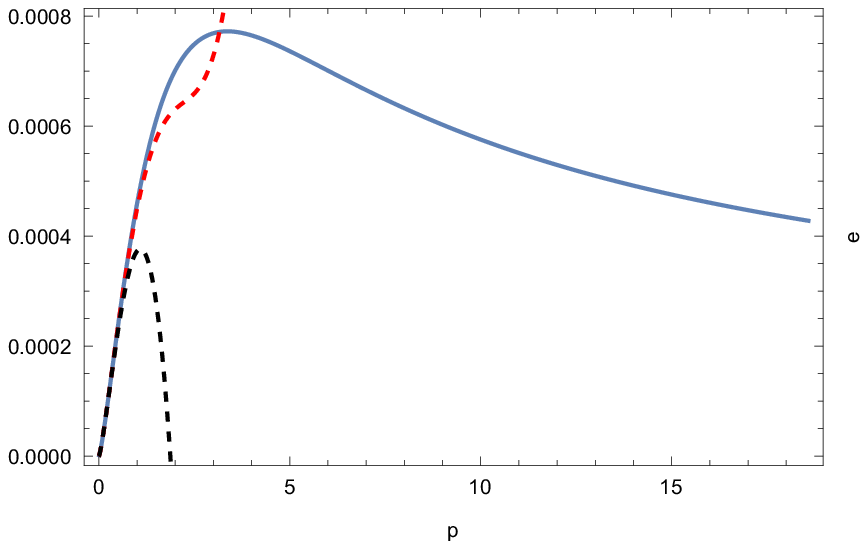}
\end{center}
  \caption{The expectation value  $\calo_\phi$
  (the left panel) and the vacuum entanglement entropy density
  $s_{ent}$ (the right panel)
  of the $\zet_2^\chi$-symmetric DFP (solid curves). The black/red
  dashed curves present the (successive) perturbative in the limit
  $\frac{\Lambda}{H}\to 0$ approximations to the observables.
} \label{osunbroken}
\end{figure}

\begin{figure}[h]
\begin{center}
\psfrag{a}[cc][][0.8][0]{$\ \  \hw_4$}
\psfrag{b}[cc][][0.8][0]{$\ \  \hw_5$}
\psfrag{c}[cc][][0.8][0]{$\ \  \hw_6$}
\psfrag{p}[cc][][1][0]{$\frac{\Lambda}{H}$}
\psfrag{w}[bb][][1][0]{$\Im[\hw]$}
\psfrag{t}[tt][][1][0]{$\Im[\hw]$}
\includegraphics[width=3in]{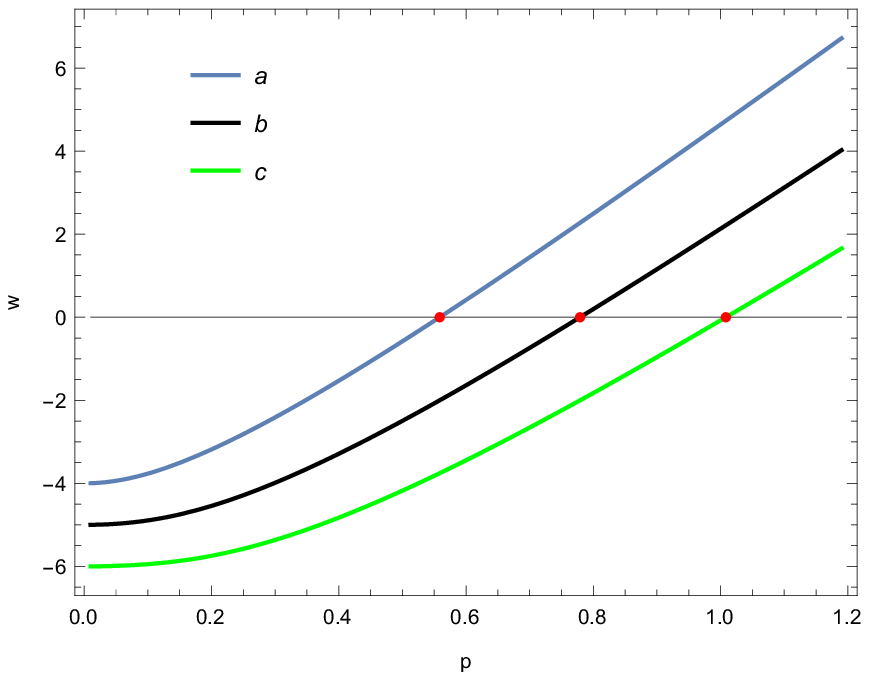}
\includegraphics[width=3in]{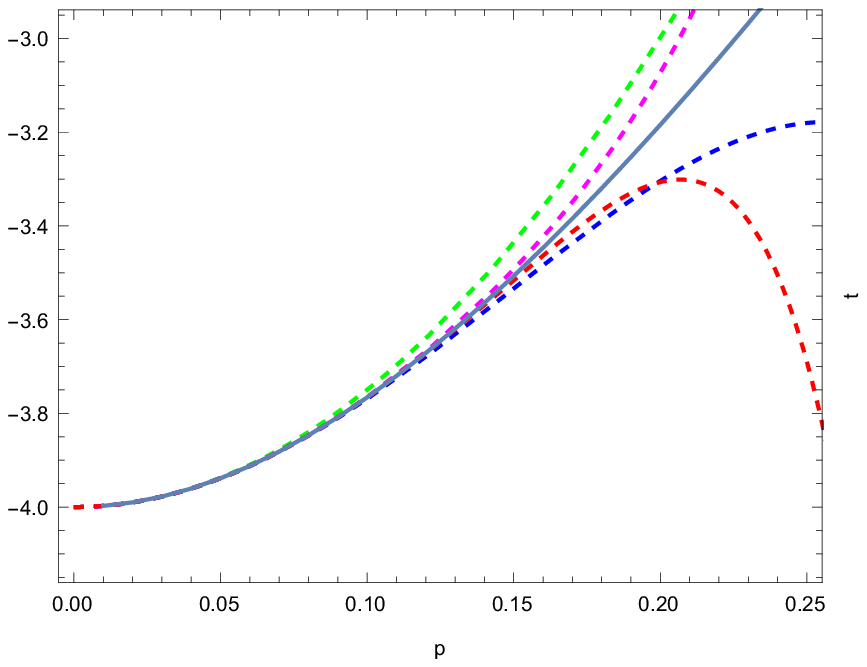}
\end{center}
  \caption{Spectra $\hw_n=\hw_n(\Lambda/H)$
  of $\zet_2^\chi$ symmetry breaking fluctuations about
  the \dfps. The red dots indicate the onset of the
  instability, see \eqref{onset}. In the limit
  $\frac \Lambda H\to 0$, the $\hw_n$ spectra can be computed
  analytically. The right panel shows $\hw_4$ (the solid curve),
  and the dashed curves are the successive perturbative approximations,
  see \eqref{pertwu4}.
} \label{wunbroken}
\end{figure}

The effective action \eqref{s4} can be consistently truncated
to a $\zet_2^\chi$-symmetric sector, \ie setting
\begin{equation}
\chi(\tau,r)\equiv 0\,.
\eqlabel{chi0}
\end{equation}
This truncation has been studied extensively in \cite{Buchel:2017lhu}.
In appendix \ref{dfpu} we present the technical details, pertinent to the
construction of the \dfps. This DFP exists for
\begin{equation}
\frac{|\Lambda|}{H}\ > 0\,.
\eqlabel{uexist}
\end{equation}
The $\zet_2^\phi$ symmetry is explicitly broken by $\Lambda\ne 0$,
the results presented are invariant under $\Lambda\leftrightarrow -\Lambda$;
we assume $\Lambda >0$.

In figs.~\ref{epunbroken} and \ref{osunbroken} we present  the results for
the energy density $\cale$, the pressure $P$,
the expectation value of the operator $\calo_\phi$, and the
VEE density of the \dfps, as one varies $\frac\Lambda H$ (the solid curves).
In the limit $\frac \Lambda H\to 0$ all the observables can be computed
analytically --- see \eqref{pertvevs} and \eqref{sentpertu}.
The red/black dashed curves represent the highest/next-to-highest
(computed) perturbative
approximations to the given observable.

As explored in \cite{Buchel:2017lhu}, \dfps is perturbatively stable
under the strict truncation \eqref{chi0}. In appendix \ref{dfpflu}
we study the spectrum of $\zet_2^\chi$ symmetry breaking fluctuations
about the \dfps. The results are collected in fig.~\ref{wunbroken}.
In the limit $\frac{\Lambda}{H}\to 0$ the spectrum of the QNMs
can be computed analytically --- see \eqref{pertwu4} for 
the $\hw_4$ mode (the solid curve in the right panel; the dashed curves
represent the successive perturbative approximations). 
While these linearized fluctuations are stable for small enough
$\frac{\Lambda}{H}$, they all eventually become unstable
(solid curves, the left panel):
\begin{equation}
\begin{split}
&\Im[\hw_4]> 0\qquad \Longrightarrow\qquad \frac{\Lambda}{H}>
p_{1,4}^{crit} \equiv 0.55867(6)\,;
\\
&\Im[\hw_5]> 0\qquad \Longrightarrow\qquad \frac{\Lambda}{H}> p_{1,5}^{crit} \equiv0.77919(5)\,;
\\
&\Im[\hw_6]> 0\qquad \Longrightarrow\qquad \frac{\Lambda}{H}> p_{1,6}^{crit} \equiv1.0082(4)\,.
\end{split}
\eqlabel{instability}
\end{equation}
The red dots indicate the onset of the instability
\begin{equation}
\Im[\hw_n]\ \bigg|_{\frac\Lambda H=p_{1,n}^{crit}}=0\,.
\eqlabel{onset}
\end{equation}

\subsection{DFP with spontaneously broken $\zet_2^\chi$ symmetry --- \dfpb}

\begin{figure}[h]
\begin{center}
\psfrag{p}[cc][][1][0]{$\frac{\Lambda}{H}$}
\psfrag{e}[bb][][1][0]{$\frac{\cale}{cH^3}$}
\psfrag{r}[tt][][1][0]{$\frac{P}{cH^3}$}
\includegraphics[width=3in]{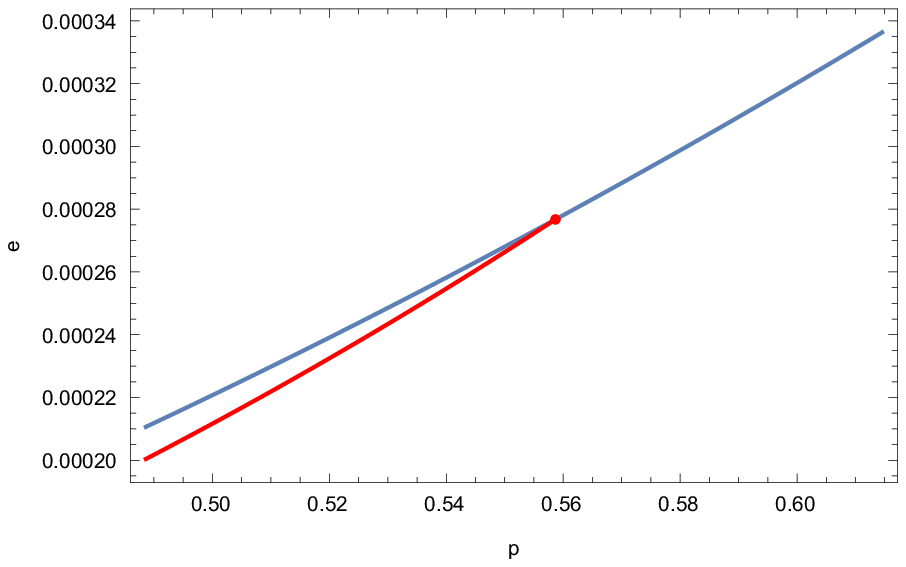}
\includegraphics[width=3in]{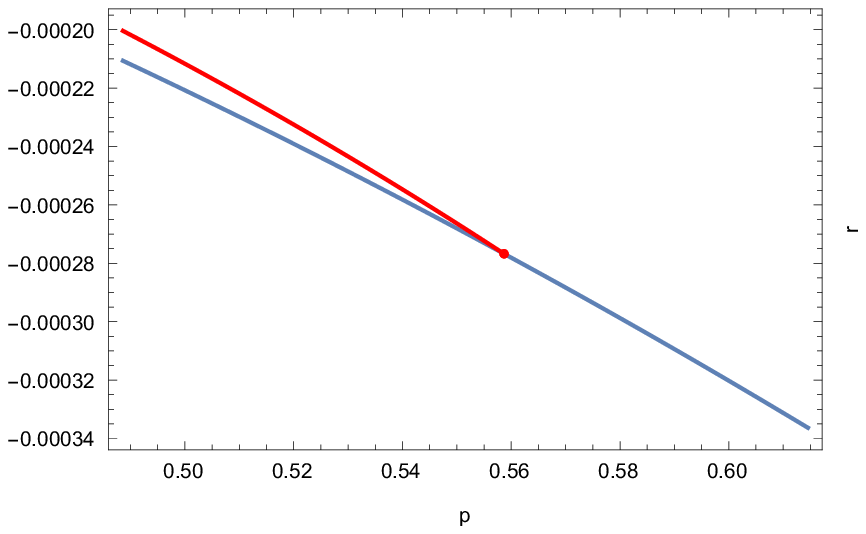}
\end{center}
  \caption{The energy density $\cale$ (the left panel) and the pressure $P$ (the right panel)
  of $\zet_2^\chi$-symmetric DFP (the blue curves) and the symmetry broken DFP
  (the red curves). \dfpb exists only above certain critical value of the Hubble constant,
  see \eqref{brange}.
} \label{epbroken}
\end{figure}

\begin{figure}[h]
\begin{center}
\psfrag{p}[cc][][1][0]{$\frac{\Lambda}{H}$}
\psfrag{q}[bb][][1][0]{$\frac{\calo_\chi}{cH^4}$}
\psfrag{e}[tt][][1][0]{$\frac{s_{ent}}{4\pi cH^2}$}
\includegraphics[width=3in]{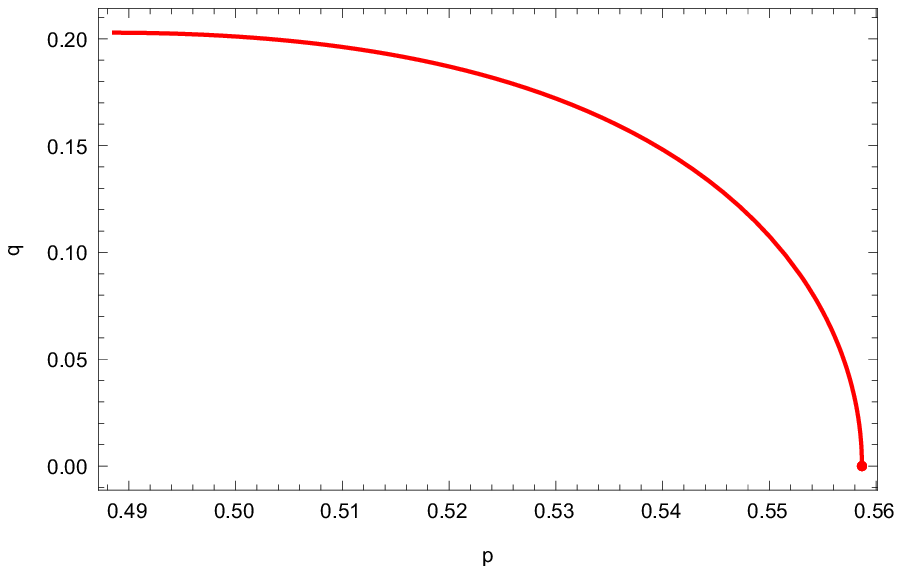}
\includegraphics[width=3in]{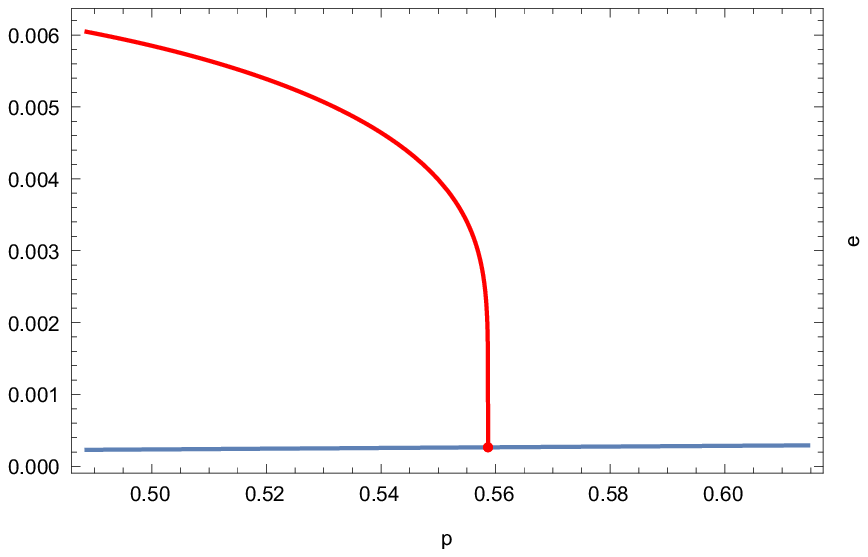}
\end{center}
  \caption{The left panel: the expectation value of the $\zet_2^\chi$ symmetry breaking
  order parameter $\calo_\chi$. The right panel: the vacuum entanglement entropies of
  the \dfps (the blue curve) and the \dfpb (the red curve). 
} \label{sbroken}
\end{figure}

\begin{figure}[h]
\begin{center}
\psfrag{p}[cc][][1][0]{${\Lambda}/{H}$}
\psfrag{q}[cc][][1][0]{${\calo_\chi}/{(cH^4)}$}
\psfrag{w}[bb][][1][0]{$\Im[\hw]$}
\psfrag{t}[tt][][1][0]{$\Im[\hw]$}
\includegraphics[width=3in]{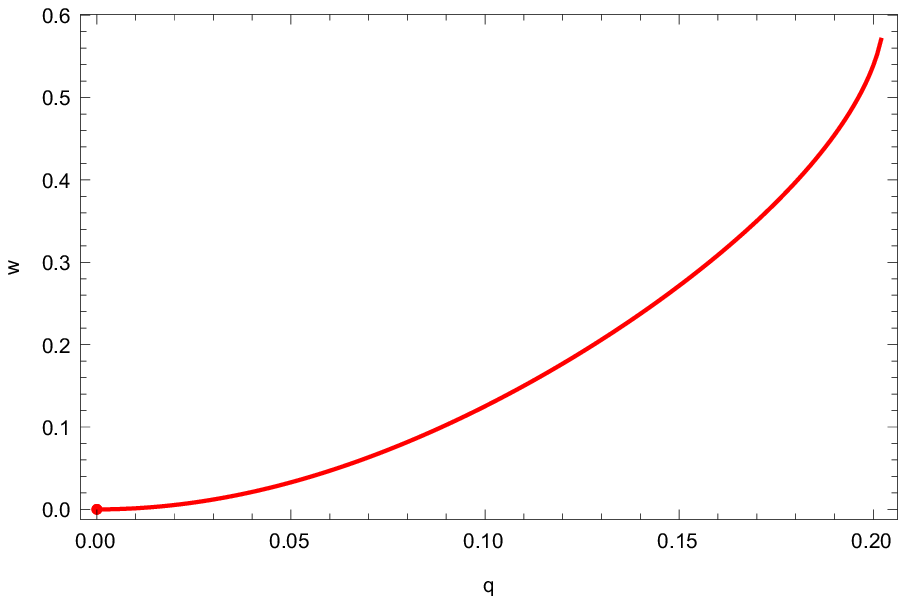}
\includegraphics[width=3in]{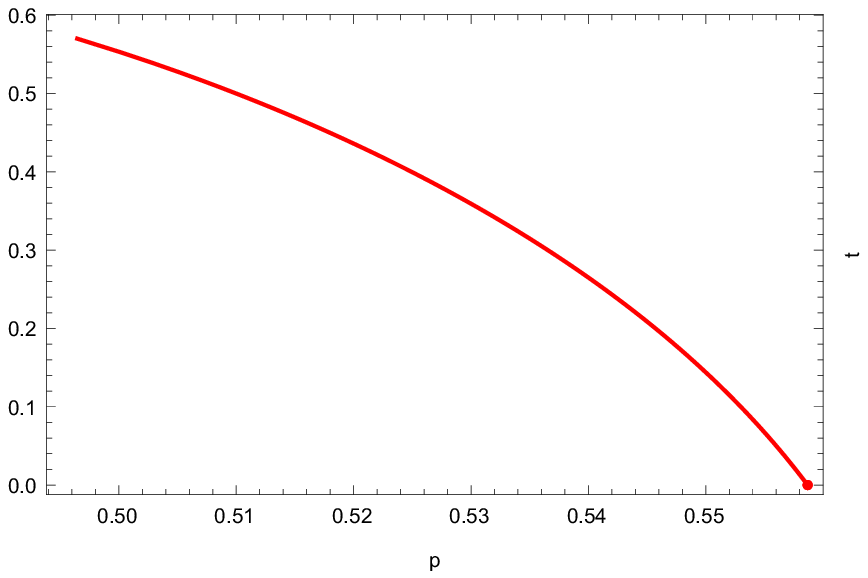}
\end{center}
  \caption{The \dfpb is perturbatively unstable. The left panel: $\hw$ of the unstable QNM
  as a function of the symmetry breaking order parameter in the \dfpb. The right panel:
  $\hw$ as a function of $\frac \Lambda H$ of \dfpb.  
} \label{wbroken}
\end{figure}

The onset of the $\zet_2^\chi$ spontaneous symmetry breaking instability
associated with the QNM $\hw_n$ about \dfps, see 
\eqref{onset}, represent a coexistence point of \dfps and
a new, symmetry broken, dynamical fixed point ${\rm{DFP}}_b^n$.   
In appendix \ref{dfpb} we collect the technical details, pertinent to the
construction of the ${\rm{DFP}}_b^n$.
We present the results
only\footnote{It is straightforward to construct higher-$n$
dynamical fixed points --- we leave the detailed analysis for the
future. 
} for ${\rm{DFP}}_b^4$, henceforth denoted
simply as \dfpb.

We constructed \dfpb  for a fairy narrow range 
\begin{equation}
\frac\Lambda H\ \in (\approx 0.4886, p_{1,4}^{crit}=0.55867(6)]\,.
\eqlabel{brange}
\end{equation}
The upper bound is set by the onset of the instability due to
$\hw_4$, see \eqref{instability}; and the lower bound is
due to the limitations of the numerical codes we used. We determined that
as $H$ increases, the Kretschmann scalar evaluated at the apparent
horizon grows --- it increases over the range \eqref{brange} by a
factor
of $\approx 66$, making the dual supergravity approximation less reliable. 
There must exist an obstruction for a \dfpb  as
$\frac\Lambda H\to 0$: in this limit the model is conformal.
Note that the symmetry broken DFP exists for $H> \frac{1}{p_{1,4}^{crit}}\ \Lambda$
--- this is similar to the DFP of the cascading gauge theory with spontaneously
broken chiral symmetry, which also
exists for $H> H_{min}\propto \Lambda$ \cite{Buchel:2019pjb}.

In fig.~\ref{epbroken} we compare the energy density and the pressure of the
\dfps (the blue curves) and the \dfpb (the red curves). While the stress-energy tensor
of the model is renormalization scheme dependent, see \eqref{vev1} and \eqref{vev2},
the comparison is meaningful in a fixed scheme: notice that the \dfpb has the lower
energy density, and the higher pressure. 

In the left panel of fig.~\ref{sbroken} we present the expectation value
of the order parameter $\calo_\chi$ for the spontaneous breaking of $\zet_2^\chi$
symmetry in the \dfpb. In the right panel we compare the VEE densities $s_{ent}$
of the \dfps (the blue curve) and the \dfpb (the red curve). Note that whenever both exist,
the VEE density of the \dfpb is the large one  --- suggesting that it, rather
than the \dfps, is the attractor of the late-time dynamics. However,
whenever the \dfps exists, see \eqref{brange}, the symmetry preserving
dynamical fixed point ( \dfps) is perturbatively stable, see fig.~\ref{wunbroken}.
This immediately implies that the \dfps must be non-perturbatively unstable,
if $\frac{\Lambda}{H}$ is in the range \eqref{brange}.
Our numerical simulations confirm that this is indeed the case.

An interesting twist to the story is the fact that the \dfpb is always
perturbatively unstable. The technical details are collected in appendix
\ref{dfpbfl}. In fig.~\ref{wbroken} we present the spectrum of the unstable
QNM in the \dfpb as a function of the order parameter $\calo_\chi$ (the left panel),
and of  $\frac{\Lambda}{H}$ (the right panel).

\subsection{Dynamics of the holographic model with linearized
$\zet_2^\chi$ sector}

\begin{figure}[h]
\begin{center}
\psfrag{t}[cc][][1][0]{$t=\tau H$}
\psfrag{m}[bb][][1][0]{$\mu$}
\psfrag{n}[tt][][1][0]{$\mu$}
\includegraphics[width=3in]{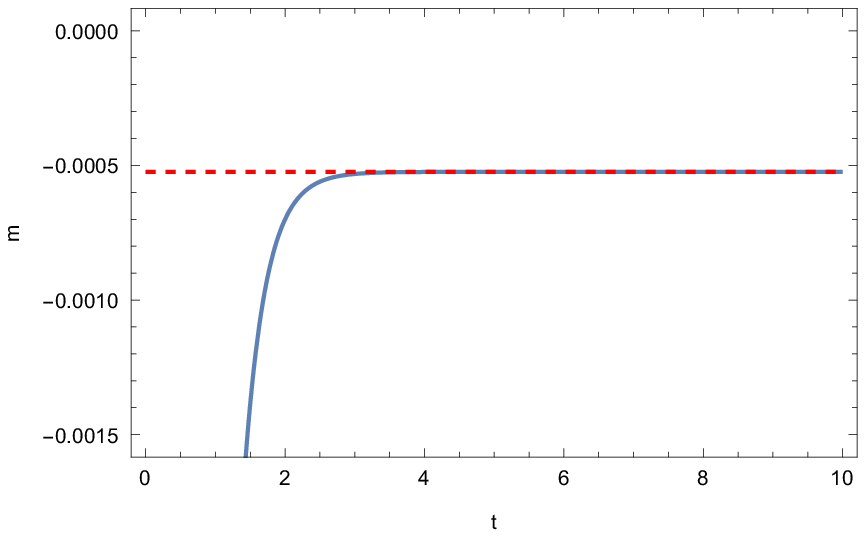}
\includegraphics[width=3in]{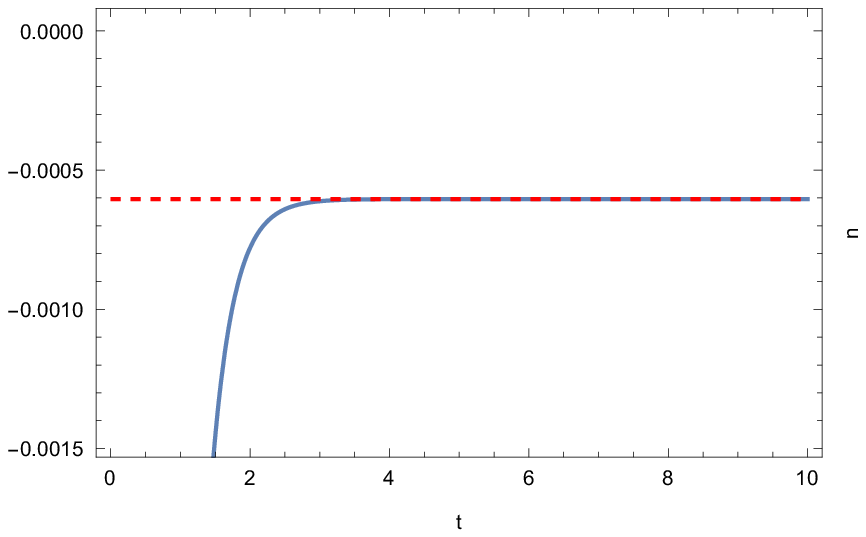}
\end{center}
  \caption{Parameter $\mu(t)$, determining the energy density $\cale$,
  see \eqref{vev1}, for  $\frac \Lambda H=p_1^<$
  (the left panel) and for  $\frac \Lambda H=p_1^>$ (the right panel),
  see \eqref{pvalues}. The dashed red lines indicate the corresponding
  \dfps attractor values. 
} \label{muuroken}
\end{figure}

\begin{figure}[h]
\begin{center}
\psfrag{t}[cc][][1][0]{$t=\tau H$}
\psfrag{p}[bb][][1][0]{$\frac{\calo_\phi}{cH^2}$}
\psfrag{o}[tt][][1][0]{$\frac{\calo_\phi}{cH^2}$}
\includegraphics[width=3in]{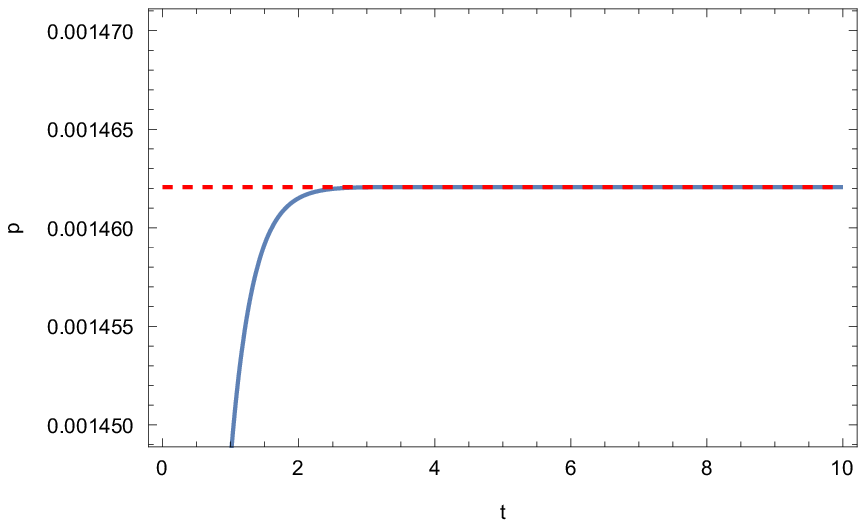}
\includegraphics[width=3in]{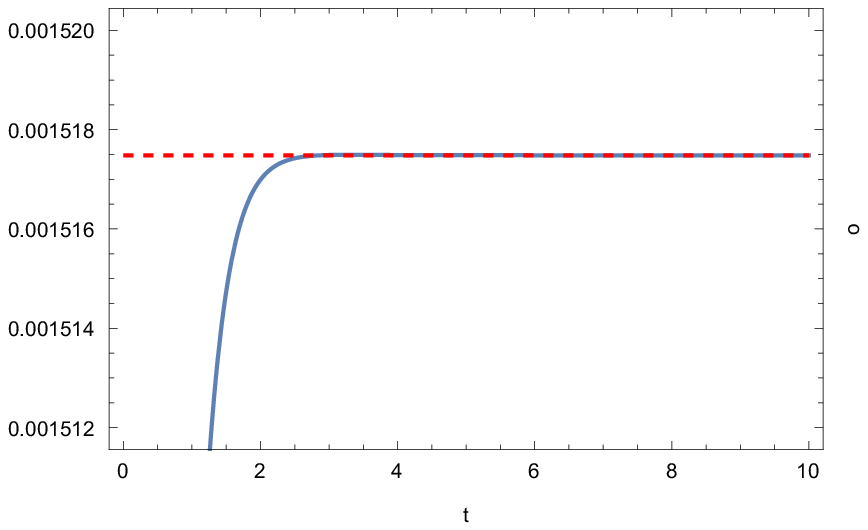}
\end{center}
  \caption{$\calo_\phi$ expectation values  for
  simulations with $\frac \Lambda H=p_1^<$
  (the left panel) and for  $\frac \Lambda H=p_1^>$ (the right panel),
  see \eqref{pvalues}.  The dashed red lines indicate the corresponding
  \dfps attractor values. 
} \label{ophiuroken}
\end{figure}

\begin{figure}[h]
\begin{center}
\psfrag{t}[cc][][1][0]{$t=\tau H$}
\psfrag{s}[bb][][1][0]{$\frac{s}{4\pi cH^2}$}
\psfrag{a}[tt][][1][0]{$\frac{s}{4\pi cH^2}$}
\includegraphics[width=3in]{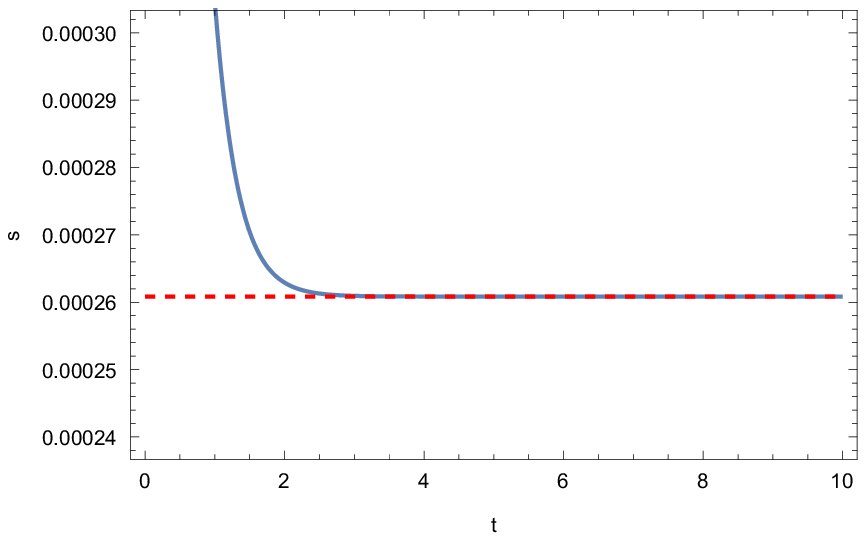}
\includegraphics[width=3in]{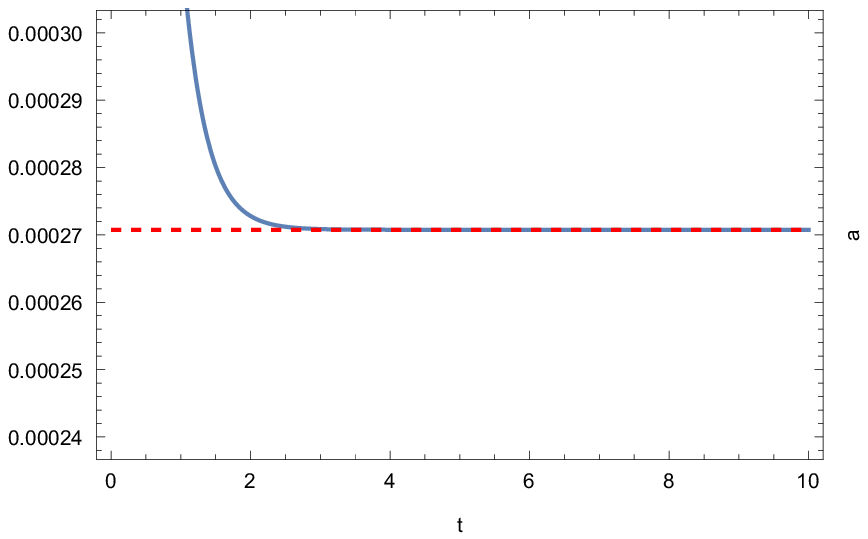}
\end{center}
  \caption{Dynamical entropy density  for
  simulations with $\frac \Lambda H=p_1^<$
  (the left panel) and for  $\frac \Lambda H=p_1^>$ (the right panel),
  see \eqref{pvalues}.  The dashed red lines indicate the corresponding
  \dfps attractor values of the vacuum entanglement entropy density $s_{ent}$. 
} \label{senturoken}
\end{figure}

\begin{figure}[h]
\begin{center}
\psfrag{t}[cc][][1][0]{$t=\tau H$}
\psfrag{w}[bb][][1][0]{$\del_t [\ln {\calo_\chi}]$}
\psfrag{q}[tt][][1][0]{$\del_t [\ln {\calo_\chi}]$}
\includegraphics[width=3in]{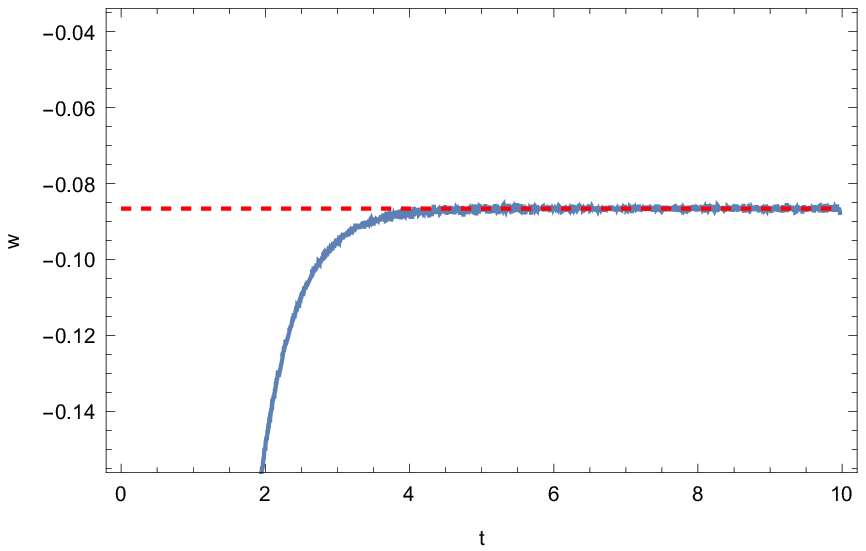}
\includegraphics[width=3in]{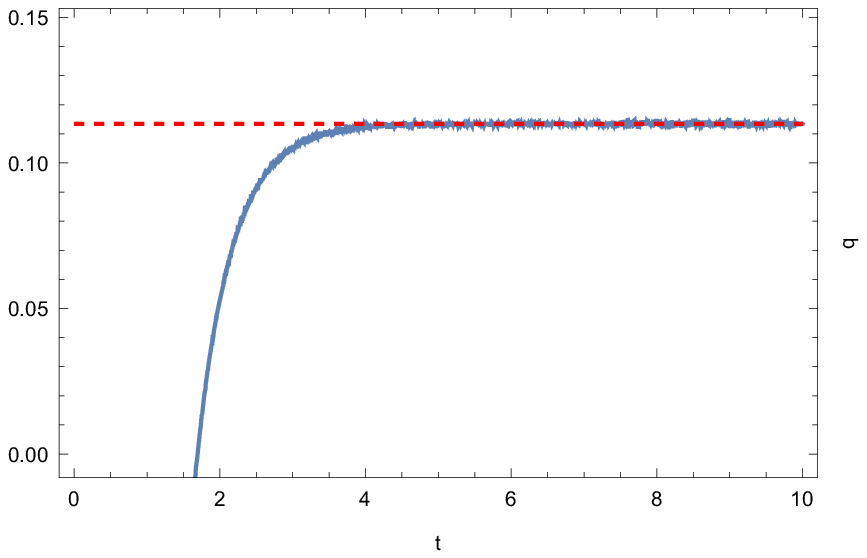}
\end{center}
  \caption{Dynamics of the $\zet_2^\chi$ symmetry breaking order parameter for
  simulations with $\frac \Lambda H=p_1^<$
  (the left panel) and for  $\frac \Lambda H=p_1^>$ (the right panel),
  see \eqref{pvalues}.  The dashed red lines indicate the expected
  late-time asymptotes, as predicted from the QNM analysis,
  see \eqref{expless} and \eqref{expmore}.
} \label{ochiuroken}
\end{figure}

Technical details used to develop the simulation code are collected in
appendix \ref{num}.
The simulations are done with time $\tau$ measured in units of the Hubble
time, see \eqref{defx}.
As a warm-up, we present here the simulation results
of our holographic model in the limit when the backreaction of the
symmetry breaking sector $\chi$, the $\call_i$ Lagrangian of the 
effective action \eqref{s4}, is neglected. The remaining $\call_{CFT}+\call_r$
sector is simulated fully
nonlinear\footnote{This was done originally in \cite{Buchel:2017lhu}.},
and the $\chi$-sector is evolved as a 'probe'. 

We use initial conditions as described in appendix \ref{initial} with
\begin{equation}
\cala_p=0.1\,,\qquad \cala_q=1\,,\qquad \mu\bigg|_{t=0}=-0.1\,.
\eqlabel{in1}
\end{equation}
Since $\chi$-dynamics is linear and in the probe approximation, the amplitude
of the initial profile $\cala_q$ is an irrelevant overall scale parameter.
The initial value of $\mu(t)$ determines the energy density of the initial holographic
state, see \eqref{vev1}. We compare simulation runs for two values of the
parameter $p_1$:
\begin{equation}
p_1\equiv \frac{\Lambda}{H}\ =\ \{\ p_1^{<}\equiv 0.55\,,\, p_1^>\equiv 0.57\ \}\,.
\eqlabel{pvalues}
\end{equation}
Note that $p_1^< < p_{1,4}^{crit}$, while  $p_1^> >  p_{1,4}^{crit}$ ---
thus we expect that simulations with $p_1=0.55$ would relax to
the corresponding \dfps with
\begin{equation}
\calo_\chi\ \propto\ e^{-i \hw_4^< \cdot t}\,,\qquad \hw_4^<=-i\ 0.086582(6)\,,
\eqlabel{expless}
\end{equation}
while those with
 $p_1=0.57$ would highlight 
the corresponding \dfps perturbative instability 
\begin{equation}
\calo_\chi\ \propto\ e^{-i \hw_4^> \cdot t}\,,\qquad \hw_4^<=+i\ 0.113428\,,
\eqlabel{expmore}
\end{equation}
for $t\gg 1$. 
The values  $\hw_4^<$ and $\hw_4^>$ are extracted from the data files used to
generate fig.~\ref{wunbroken}.

Figs.~\ref{muuroken}-\ref{senturoken} show that the \dfps is indeed the attractor of the
late-time dynamics at $\tau H\gg1$ in the $\zet_2^\chi$-symmetric sector,
for both of the select values of $p_1$ in \eqref{pvalues}. There are no
instabilities in this sector, as established in \cite{Buchel:2017lhu}.
The solid curves represent the time-series of the corresponding observables
(extracted from the FORTRAN simulation code), and the dashed red lines
indicate the attractor values of these observables (extracted from the
Wolfram Mathematica codes for DFPs).

Fig.~\ref{ochiuroken} presents the evolution of the $\zet_2^\chi$ symmetry breaking
order parameter $\calo_\chi$ for both of the select values of $p_1$ in \eqref{pvalues},
in the probe approximation (the solid curves). The late-time dynamics of the order parameter 
agrees with the QNM analysis \eqref{expless} and \eqref{expmore}, represented by
the dashed red lines.

\subsection{\dfps or \dfpb?}


\begin{figure}[h]
\begin{center}
\psfrag{a}[cc][][0.6][0]{$\ \cala_q^{(1)}$}
\psfrag{b}[cc][][0.6][0]{$\ \cala_q^{(2)}$}
\psfrag{c}[cc][][0.6][0]{$\ \cala_q^{(3)}$}
\psfrag{d}[cc][][0.6][0]{$\ \cala_q^{(4)}$}
\psfrag{e}[cc][][0.6][0]{$\ \cala_q^{(5)}$}
\psfrag{t}[cc][][1][0]{$t=\tau H$}
\psfrag{q}[bb][][1][0]{$\frac{\calo_\chi}{cH^4}$}
\psfrag{p}[tt][][1][0]{$\frac{\calo_\phi}{cH^2}$}
\includegraphics[width=3in]{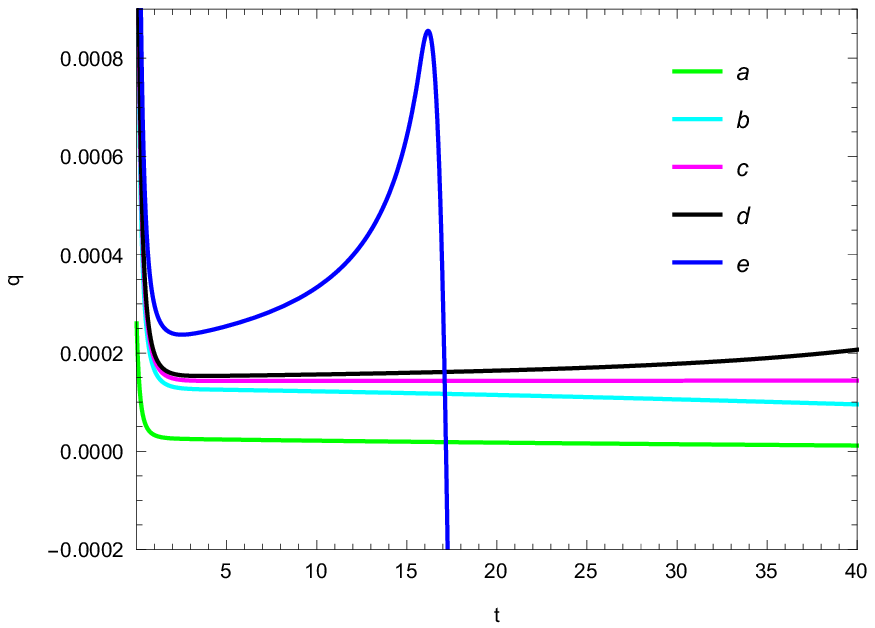}
\includegraphics[width=3in]{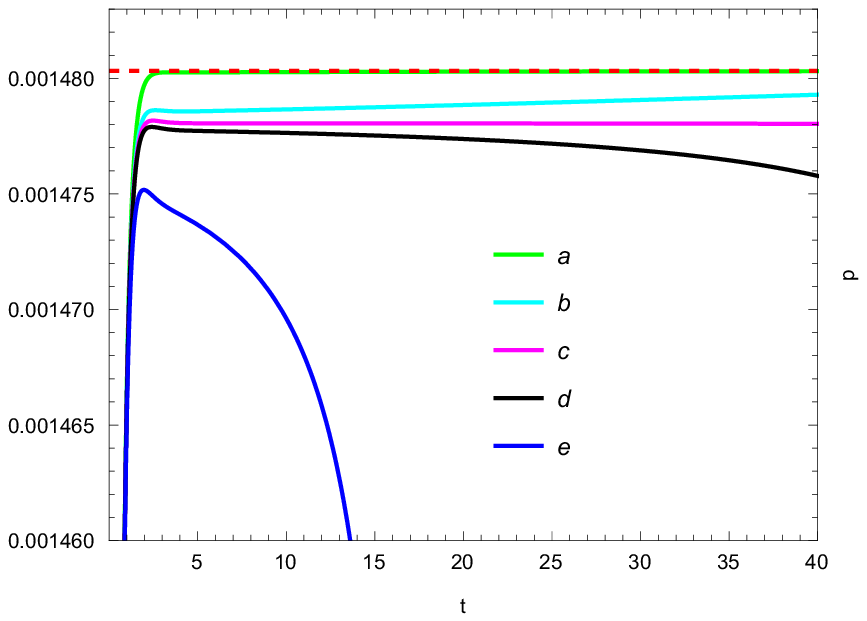}
\end{center}
  \caption{Evolution of expectation values of $\calo_\chi$ (the left panel)
  and $\calo_\phi$ (the right panel) for  simulations at $\frac \Lambda H=p_1^*$
  \eqref{pqdef}, when both the \dfps and the \dfpb exist. Initial conditions
for the amplitude of the $\zet_2^\chi$ symmetry breaking order parameter $\cala_q$,
see \eqref{initcond},  with
  $\cala_q^{(i)}< \cala_q^{crit}\approx \cala_q^{(i)}$, see \eqref{aqset}, evolve
  to the \dfps; while those with  $\cala_q^{(i)}> \cala_q^{crit}$ eventually evolve to a
  naked singularity. The fine-tuned $\cala_q^{(3)}$ initial condition approaches the
  perturbatively unstable \dfpb at late times. 
} \label{aqqp}
\end{figure}

\begin{figure}[h]
\begin{center}
\psfrag{a}[cc][][0.6][0]{$\ \cala_q^{(1)}$}
\psfrag{b}[cc][][0.6][0]{$\ \cala_q^{(2)}$}
\psfrag{c}[cc][][0.6][0]{$\ \cala_q^{(3)}$}
\psfrag{d}[cc][][0.6][0]{$\ \cala_q^{(4)}$}
\psfrag{e}[cc][][0.6][0]{$\ \cala_q^{(5)}$}
\psfrag{t}[cc][][1][0]{$t=\tau H$}
\psfrag{q}[bb][][1][0]{$\del_t [\ln \calo_\chi]$}
\psfrag{k}[tt][][1][0]{$\ln\calk_{AH}$}
\includegraphics[width=3in]{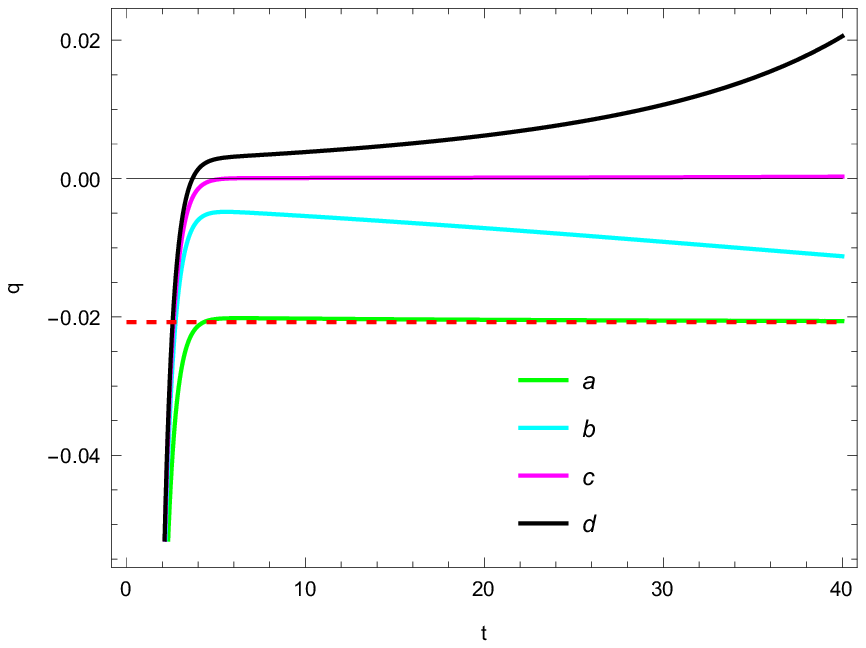}
\includegraphics[width=3in]{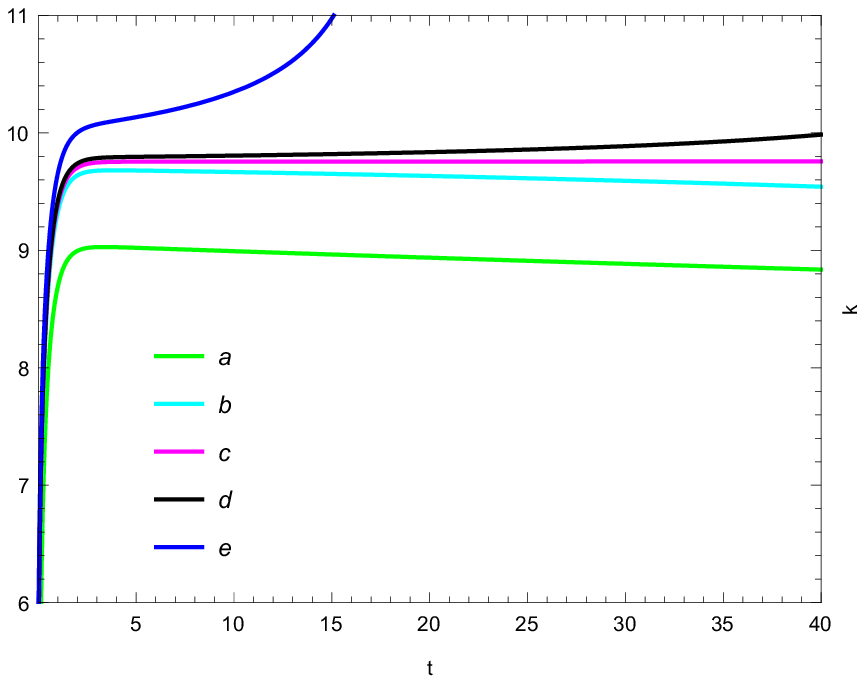}
\end{center}
  \caption{The growth rate  of the expectation value  of $\calo_\chi$ (the left panel),
and the  Kretschmann scalar $\calk$ \eqref{defk},
evaluated at the apparent horizon 
(the right panel) for  simulations at $\frac \Lambda H=p_1^*$
  \eqref{pqdef}, when both the \dfps and the \dfpb exist. The small amplitude
  initial condition $\cala_q^{(1)}$ (the green curve) approaches the \dfps with the appropriate
  decay rate, determined by the \dfps QNM with $\Im[\hw_{4}(p_1^*)]$ (the dashed red line).
  The fine-tuned initial condition  $\cala_q^{(3)}$ (the magenta curve) approaches
  at late times the \dfpb --- it has (close to) vanishing growth rate.
  Initial configurations with $\cala_q> \cala_q^{crit}$ result in uncontrollable growth
  of the  Kretschmann scalar at the apparent horizon (the black and the blue curves)
  at late times.
} \label{aqqk}
\end{figure}

In this section we present results of the simulations with initial conditions
as described in appendix \ref{initial} with
\begin{equation}
\cala_p=0.1\,,\qquad \mu\bigg|_{t=0}=-0.1\,,
\eqlabel{in2}
\end{equation}
and a set of amplitudes $\cala_q^{(i)}$ of the $\zet_2^\chi$ symmetry breaking
initial conditions:
\begin{equation}
\cala_q^{(i)}\ =\ \{\ \underbrace{0.1}_{i=1}\,,\, 0.5\,,\, \underbrace{0.564}_{i=3}\,,\, 0.6\,,\, \underbrace{0.9}_{i=5}\ \}\,.
\eqlabel{aqset}
\end{equation}
We perform simulations at
\begin{equation}
\frac{\Lambda}{H}=p_1^*=0.5566\ <\ p_{1,4}^{crit}\,.
\eqlabel{pqdef} 
\end{equation}
Note that at select value of $p_1^*$  both the \dfps and the \dfpb exist:  
\nxt the \dfps is perturbatively stable;
\nxt the \dfpb is perturbatively unstable;
\nxt however,
\begin{equation}
s_{ent}({\rm DFP}_b)\ > s_{ent}({\rm DFP}_s)\,.
\eqlabel{scompare}
\end{equation}
Given above, we expect that there is a critical amplitude $\cala_q^{crit}$, such that
initial configurations with $\cala_q< \cala_q^{crit}$ evolve to the \dfps, while configurations
with $\cala_q> \cala_q^{crit}$ would evolve to a naked singularity.  $\cala_q> \cala_q^{crit}$
initial conditions can not evolve to the \dfpb, as it is unstable,
neither can they evolve to the \dfps since a more entropic configuration is available.
This is indeed what we find, with
\begin{equation}
\cala_q^{crit}\ \approx \cala_q^{(3)}\,.
\eqlabel{defaqcrit}
\end{equation}

In fig.~\ref{aqqp} we present the evolution of $\calo_{\chi}$ (the left panel) and 
$\calo_\phi$ (the right panel) for select values of $\cala_q$, see \eqref{aqset}.
For small amplitudes, $\cala_q^{(1)}$ and $\cala_q^{(2)}$,
the late-time attractor remains the \dfps --- note in particular that for
$\cala_q^{(1)}$ (the green curves), rather quickly, the expectation value of $\calo_{\chi}$
decays to zero, restoring $\zet_2^\chi$ symmetry, while the expectation value of
$\calo_\phi$ approaches the corresponding attractor value of the \dfps (the dashed red line).
Initial configurations  $\cala_q^{(4)}$ and $\cala_q^{(5)}$ result in uncontrollable growth of
the expectation values, more pronounced for $\cala_q^{(5)}$ (the blue curves).
The fine-tuned initial condition with $\cala_q^{(3)}$ evolves (close) to the \dfpb
(the magenta curves).

In fig.~\ref{aqqk} we extract the growth rate of $\calo_{\chi}$ (the left panel),
and present the evolution of the Kretschmann scalar $\calk$ \eqref{defk},
evaluated at the apparent horizon. Initial condition $\cala_q^{(1)}$ quickly
approaches the decay rate set by the corresponding \dfps (the red dashed line).
Since the fine-tuned initial condition $\cala_q^{(3)}$ evolves to the \dfpb,
the late-time growth rate of $\calo_{\chi}$ is close to zero (the magenta curve).
The Kretschmann scalar $\calk_{AH}$ grows (apparently unboundedly) at late times for
initial conditions with $\cala_q> \cala_q^{crit}$ --- the growth is the faster,
the larger is the initial amplitude of the $\zet_2^\chi$ symmetry breaking.

\subsection{DFP does not always exist}


\begin{figure}[h]
\begin{center}
\psfrag{t}[cc][][1][0]{$t=\tau H$}
\psfrag{q}[bb][][1][0]{$\ln\frac{\calo_\chi}{cH^4}$}
\psfrag{p}[tt][][1][0]{$\frac{\calo_\phi}{cH^2}$}
\includegraphics[width=3in]{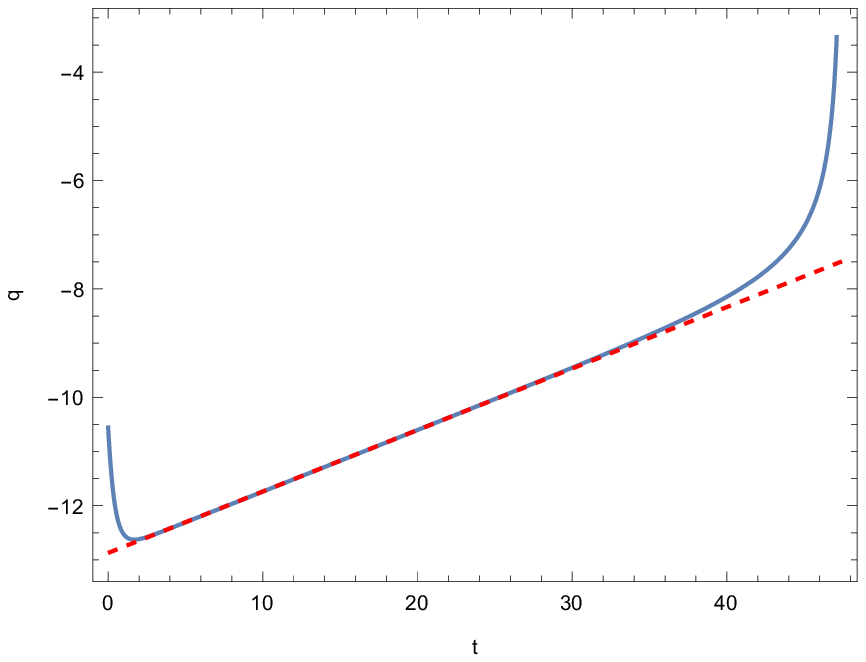}
\includegraphics[width=3in]{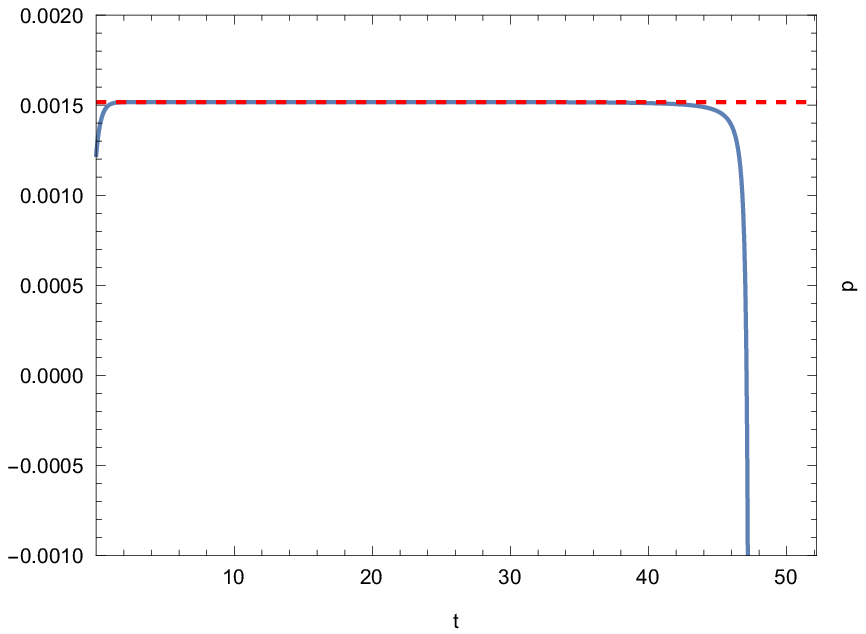}
\end{center}
  \caption{Evolution of expectation values of $\calo_\chi$ (the left panel)
  and $\calo_\phi$ (the right panel) for a simulation at $\frac \Lambda H=\tp_1$
  \eqref{pqdef2}, when only the \dfps exists. Given that the initial amplitude
  in the $\zet_2^\chi$ symmetry broken sector is small, see \eqref{in3},
  its evolution is probe-like on the \dfps characterized by \eqref{pqdef2}.
  At late times the nonlinearities in the gravitational
  scalar sector grow without bounds.
} \label{chiphi057}
\end{figure}

\begin{figure}[h]
\begin{center}
\psfrag{t}[cc][][1][0]{$t=\tau H$}
\psfrag{s}[bb][][1][0]{$\ln \frac{s}{4\pi cH^2}$}
\psfrag{k}[tt][][1][0]{$\ln\calk_{AH}$}
\includegraphics[width=3in]{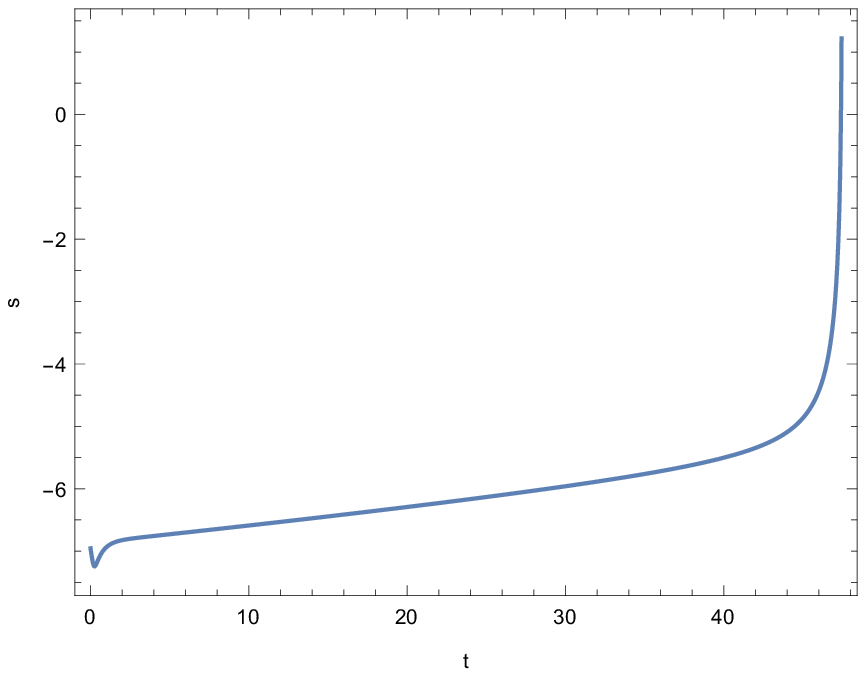}
\includegraphics[width=3in]{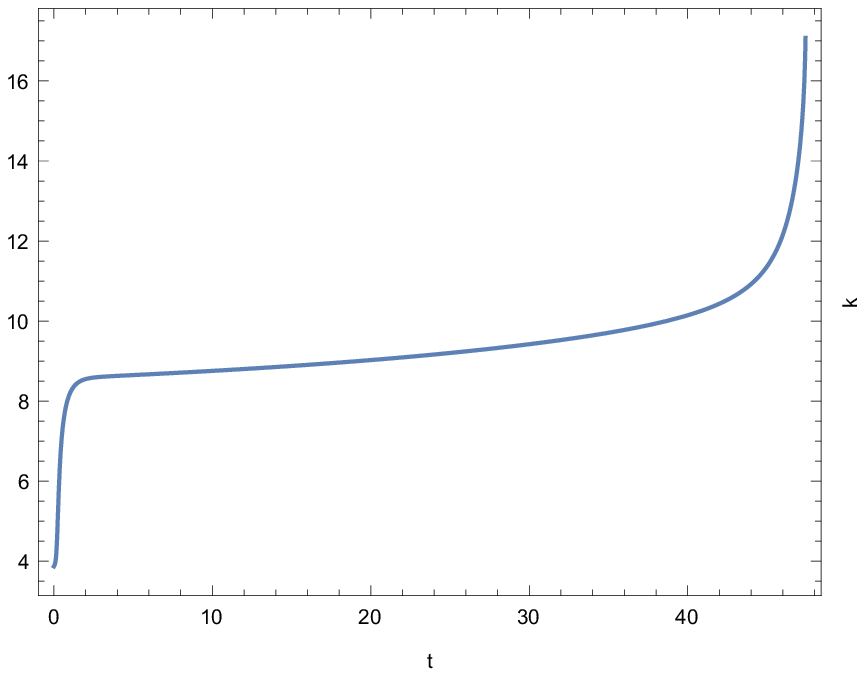}
\end{center}
  \caption{The evolution of the entropy density $s$ (the left panel),
and the  Kretschmann scalar $\calk$ \eqref{defk},
evaluated at the apparent horizon 
(the right panel) for a simulation at $\frac \Lambda H=\tp_1$
  \eqref{pqdef2}, when only the \dfps  exists. Since the \dfps is
  unstable at \eqref{pqdef2}, both quantities grow without bounds
  at late times. 
} \label{chik057}
\end{figure}

Dynamics described in the previous section
is an example when a DFP is absent for
certain initial states (those with $\cala_q > \cala_q^{crit}$),
but is an attractor for some other initial configurations
(those with $\cala_q < \cala_q^{crit}$).
In this section we present results of the simulations with initial conditions
as described in appendix \ref{initial} with
\begin{equation}
\cala_p=0.1\,,\qquad \cala_q=0.01\,,\qquad \mu\bigg|_{t=0}=-0.1\,.
\eqlabel{in3}
\end{equation}
We perform simulations at
\begin{equation}
\frac{\Lambda}{H}=\tp_1=0.57\ >\ p_{1,4}^{crit}\,.
\eqlabel{pqdef2} 
\end{equation}
Note that at select value of $\tp_1$  only the \dfps exists,
however, it is perturbatively unstable to $\zet_2^\chi$ symmetry breaking
fluctuations. Thus we expect that, no matter how small $\cala_q$  is,
the configuration would evolve to a naked singularity, \ie this is an example
when  a DFP does not exist, no matter the initial state.

In fig.~\ref{chiphi057} we present the evolution of $\calo_{\chi}$ (the left panel) and 
$\calo_\phi$ (the right panel). Note that the benefit of the small amplitude
$\cala_q$ in \eqref{in3} is that the $\zet_{2}^\chi$-symmetric sector of the model
approaches the \dfps (within several Hubble times $t\sim 1$), while the
$\zet_{2}^\chi$-symmetry breaking sector effectively evolves as a probe
till $t\sim 35$. For yet larger values of $t$, the nonlinear effects drive
the evolution of the geometry to a naked singularity. The red dashed line
in the left panel presents the prediction for the growth of $\calo_\chi$
on top of \dfps specified by \eqref{pqdef2}, driven by the QNM with $\hw_4(\tp_1)$. 
The red dashed line in the right panel presents the expectation value of
$\calo_\phi$ in \dfps specified by \eqref{pqdef2}.

In fig.~\ref{chik057} we confirm that the instability of the \dfps, and the absence of
any other DFP at $\frac{\Lambda}{H}=\tp_1$, results in unbounded growth of
the entropy density $s$ (the left panel) and the  Kretschmann scalar
evaluated at the apparent horizon $\calk_{AH}$  at late times.

\subsection{Dynamics of \dfpb}


\begin{figure}[h]
\begin{center}
\psfrag{t}[cc][][1][0]{$t=\tau H$}
\psfrag{p}[bb][][1][0]{$\frac{\calo_\phi}{cH^2}$}
\psfrag{d}[tt][][1][0]{$\del_t \left[\ln ({\calo_\phi}-\calo_\phi^{DFP})\right]$}
\includegraphics[width=3in]{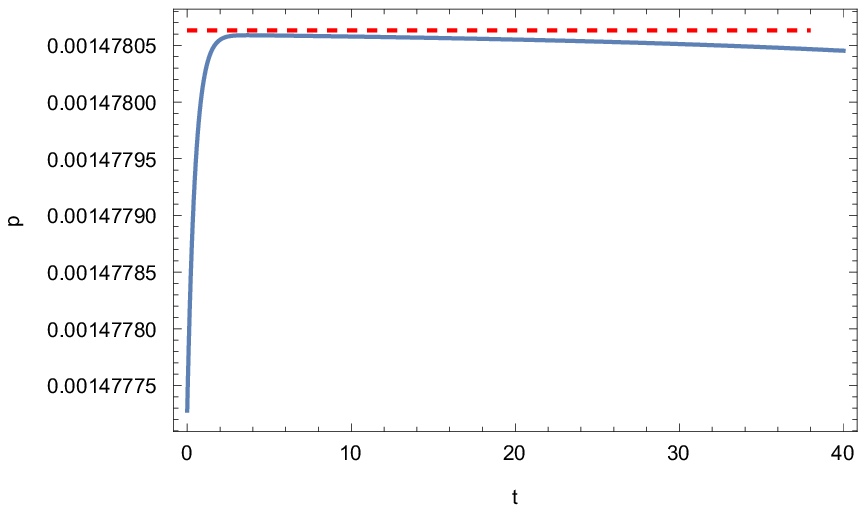}
\includegraphics[width=3in]{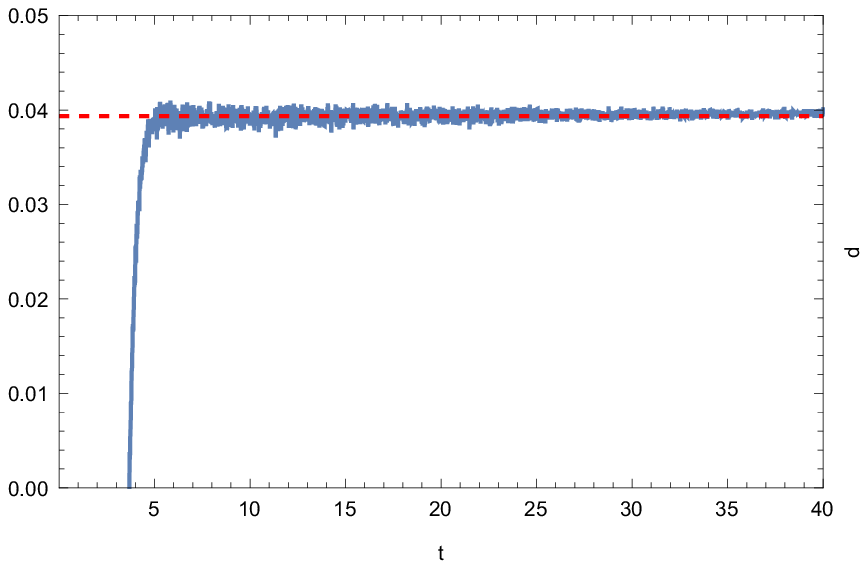}
\end{center}
  \caption{Evolution of $\calo_\phi$ with the initial configuration
  set as the \dfpb at $\frac{\Lambda}{H}=p_1^*$, see \eqref{pqdef}.
  The instability of the \dfpb causes the noise introduced by the initial conditions
  to grow at the rate predicted by the unstable quasinormal mode of the \dfpb as
  $\Im[\hw(p_1^*)]$ (the dashed red line in the right panel). 
} \label{pbroken}
\end{figure}

\begin{figure}[h]
\begin{center}
\psfrag{t}[cc][][1][0]{$t=\tau H$}
\psfrag{q}[bb][][1][0]{$\frac{\calo_\chi}{cH^4}$}
\psfrag{d}[tt][][1][0]{$\del_t \left[\ln ({\calo_\chi}-\calo_\chi^{DFP})\right]$}
\includegraphics[width=3in]{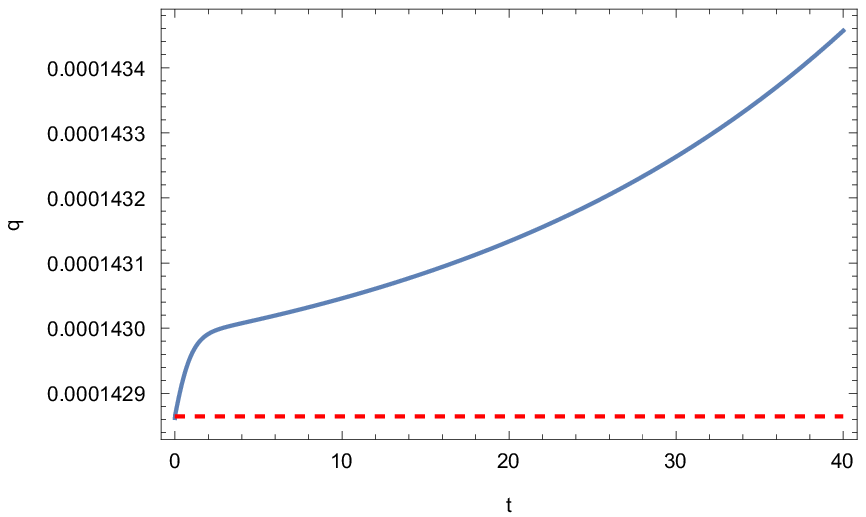}
\includegraphics[width=3in]{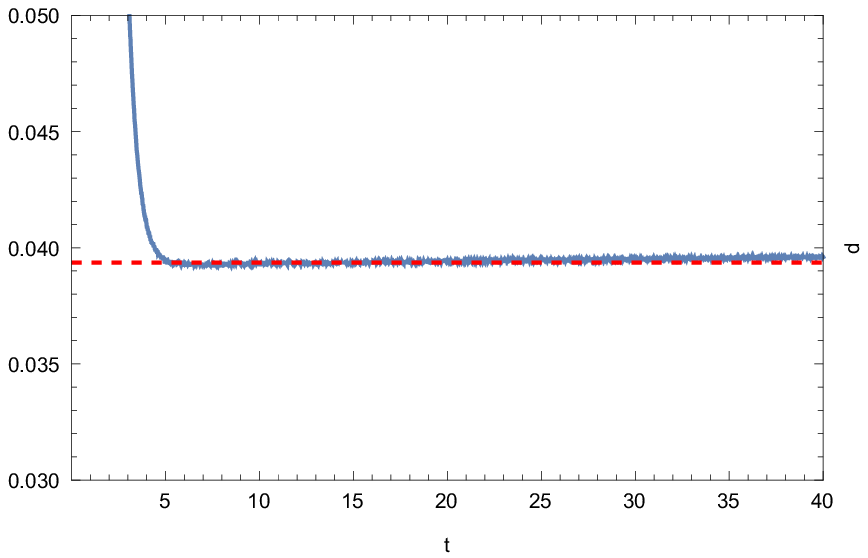}
\end{center}
  \caption{Evolution of $\calo_\chi$ with the initial configuration
  set as the \dfpb at $\frac{\Lambda}{H}=p_1^*$, see \eqref{pqdef}.
  The instability of the \dfpb causes the noise introduced by the initial conditions
  to grow at the rate predicted by the unstable quasinormal mode of the \dfpb as
  $\Im[\hw(p_1^*)]$ (the dashed red line in the right panel). 
} \label{qbroken}
\end{figure}

In this section we present results of the simulation when the initial condition is,
to a very good approximation, the \dfpb --- see appendix \ref{initialb} for the technical
details. We choose the \dfpb at \eqref{pqdef}. Since the \dfpb is perturbatively unstable,
the numerical noise introduced transcribing the input from Wolfram Mathematica to the simulation
code triggers the instability. The growth rate of the initial noise is controlled
by the unstable QNM of the \dfpb with $\Im[\hw(p_1^*)]$. This is indeed what we find:
in figs.~\ref{pbroken} and \ref{qbroken} we present the evolution of $\calo_{\phi}$
and $\calo_{\chi}$ (the solid curves). The dashed red lines indicate predictions
of the corresponding values from the \dfpb at \eqref{pqdef}.

\section{Conclusions}\label{conclusions}

In this paper we proposed a formal concept of a
dynamical fixed point (DFP) of a macroscopic system: the main distinction from the
thermal equilibrium is the condition that the entropy current divergence of the system
at late times is strictly positive. The clear shortcoming (with respect to
practical applications) is the necessity of the precise definition of the entropy current
of a system far-from-equilibrium. In strongly coupled systems, which have a dual holographic
description, the candidate entropy current can be constructed from the
gravitational entropy density of the apparent horizon.

There are multiple avenues for future studies:
\nxt regarding  the 'big questions',
\begin{itemize}
\item We studied DFPs of a holographic system driven by the background
space-time expansion. Can one define interesting protocols for time-dependence
of coupling constants, masses, external magnetic fields of a holographic model,
which result in DFPs?
\item What are the interesting plasma hydrodynamic flows that do not
have thermal equilibrium as a late-time attractor?
\item How universal are DFPs, and is there a universal replacement for
hydrodynamics, as a theory describing the approach to a DFP?
\item Can DFPs be studied for theories at weak coupling? Are there phenomenological
applications to cosmology?
\item Related to above, can examples of DFPs be constructed within kinetic theory?
\end{itemize}
$\ $
\nxt regarding the holographic toy model discussed here, and its generalizations,
\begin{itemize}
\item What are the properties of the \dfpb, triggered by the instabilities of
the QNMs with $\hw_n$ for $n>4$?
\item What is the landscape of DFPs of the model defined on compact spatial manifolds?
\item What is the role of supersymmetry and the unboundedness of the scalar
potentials in the dual gravitation effective actions\footnote{The work
\cite{Buchel:2017map} could be a useful resource.} on DFPs?
\item Analysis of the cascading gauge theory in de Sitter space-time
\cite{Buchel:2019pjb} {\it is} the analysis of its DFPs. Whether these
DFPs are stable or not, and when are they the attractors of the
late-time evolution remains to be analyzed.
\end{itemize}

\section*{Acknowledgments}
This research is supported in part by Perimeter Institute for
Theoretical Physics.  Research at Perimeter Institute is supported in
part by the Government of Canada through the Department of Innovation,
Science and Economic Development Canada and by the Province of Ontario
through the Ministry of Colleges and Universities. This work was
further supported by NSERC through the Discovery Grants program.

\appendix

\section{Holographic EOMs,  the boundary asymptotics and the renormalization}\label{eoms}

Einstein equations from \eqref{s4} define the following evolution equations of motion:
\begin{equation}
\begin{split}
&0=(d_+\Sigma)'+d_+\Sigma\ \left(\ln\Sigma\right)'-\frac 32 \Sigma-\frac 14\Sigma\left(\phi^2-2\chi^2
-g\phi^2\chi^2\right) \;,\\
&0=(d_+\phi)'+d_+\phi\ \left(\ln\Sigma\right)'+\frac{d_+\Sigma}{\Sigma}\ \phi'
+\phi\left(1-g \chi^2\right)\;,\\
&0=(d_+\chi)'+d_+\chi\ \left(\ln\Sigma\right)'+\frac{d_+\Sigma}{\Sigma}\ \chi'
-\chi\left(2+g \phi^2\right)\;,\\
&0=A''-2\frac{d_+\Sigma}{\Sigma^2}\ \Sigma'+\frac 12 d_+\phi\ \phi'+\frac 12
d_+\chi\ \chi' \;,
\end{split}
\eqlabel{evolveoms}
\end{equation}
together with the Hamiltonian constraint equation:
\begin{equation}
\begin{split}
&0=\Sigma''+\frac 14\Sigma\left( (\phi')^2+(\chi')^2\right) 
\;,\\
\end{split}
\eqlabel{coneoms1}
\end{equation}
and the momentum constraint equation:
\begin{equation}
\begin{split}
&0 =d_+^2\Sigma-2 A (d_+\Sigma)'-\frac{d_+\Sigma}{\Sigma^2}\
\left(A\Sigma^2\right)' \\
&  +\frac 14\Sigma \left( (d_+\phi)^2+(d_+\chi)^2 
 +2A\left(6+\phi^2-2\chi^2-g\phi^2\chi^2 \right) \right)\;, 
\end{split}
\eqlabel{coneoms2}
\end{equation}
where $'\equiv\del_r$ and $d_+\equiv \del_\tau+A\ \del_r$.
The constraint equations are preserved by the evolution equations provided they
are satisfied at a given time-like 
surface 
--- which in our case is the AdS boundary.  

It is convenient to compactify the gravitational dual spatial domain, introducing a new
dimensionless radial coordinate $x$, and measuring time in Hubble units $t$:
\begin{equation}
x\equiv \frac{H}{r}\,,\qquad t\equiv H\tau\,.
\eqlabel{defx}
\end{equation}
Further introducing
\begin{equation}
\Sigma(t,x)\equiv H\ \exp(t)\ \sigma(t,x)\,,\qquad A(t,x)\equiv H^2\ a(t,x)\,,
\eqlabel{defxt}
\end{equation}
the $H$ dependence drops out of \eqref{evolveoms} and \eqref{coneoms2}:
\begin{equation}
\begin{split}
&0=(d_+\sigma)'+d_+\sigma\ \left(\ln\sigma\right)'
+2\sigma'+\frac {3\sigma}{2x^2}+\frac {\sigma}{4x^2}\left(\phi^2-2\chi^2
-g\phi^2\chi^2\right) \;,\\
&0=(d_+\phi)'+d_+\phi\ \left(\ln\sigma\right)'+\left(\frac{d_+\sigma}{\sigma}+1\right)\ \phi'
-\frac{\phi}{x^2}\left(1-g \chi^2\right)\;,\\
&0=(d_+\chi)'+d_+\chi\ \left(\ln\sigma\right)'+\left(\frac{d_+\sigma}{\sigma}+1\right)\ \chi'
+\frac{\chi}{x^2}\left(2+g \phi^2\right)\;,\\
&0=a''+\frac 2x\ a'+\frac{2\sigma'}{x^2\sigma}\ \left(\frac{d_+\sigma}{\sigma}+1\right)-\frac {\phi'}{2x^2}\ d_+\phi-\frac {\chi'}{2x^2}\
d_+\chi\;,
\end{split}
\eqlabel{evolveomsx}
\end{equation}
and
\begin{equation}
\begin{split}
&0=\sigma''+\frac 2x\ \sigma'+\frac 14\sigma\left( (\phi')^2+(\chi')^2\right) 
\;,\\
&0 =d_+^2\sigma+2x^2a\ (d_+\sigma)'+d_+\sigma\
\left(\frac{x^2}{\sigma^2}\ \left(a\sigma^2\right)'+2\right)
+\frac{x^2}{\sigma^3}\ (a\sigma^4)'\\
&  +\frac 14\sigma \left( (d_+\phi)^2+(d_+\chi)^2 
 +2a\left(6+\phi^2-2\chi^2-g\phi^2\chi^2 \right)+4 \right)\;, 
\end{split}
\eqlabel{coneomsx}
\end{equation}
where now $'\equiv\del_x$ and $d_+\equiv \del_\tau-a\ x^2\ \del_x$.

The general asymptotic boundary  ($x\to 0$) solution  of the equations of motion, given by
\begin{equation} 
\begin{split}
\sigma=&\frac 1x+\lambda(t)-\frac 18 p_1^2\ x +\calo\left(x^2\right)
\;,\\
a=&\frac{1}{2x^2}+\frac{\lambda(t)-1}{x} -\frac 18 p_1^2+\frac 12\lambda(t)^2-\lambda
-\dot\lambda(t)  \\
&  + \left(\mu(t)-\frac 14 p_1 p_2(t)-\frac 14 p_1^2\lambda(t)+\frac 16 p_1^2\right)\ x
+\calo\left(x^2\right)  \;, \\
\phi=&{p_1}\ x+{p_2(t)}\ {x^2}+\calo\left(x^3\right)\, ,  \\
\chi=&{q_4(t)}\ {x^4}+\calo\left({x^5}\right) \, ,
\end{split}
\eqlabel{basym}
\end{equation}
with $\dot{}\equiv \del_t$, is characterized by a single constant $p_1$, and four
dynamical variables  $\{p_2(t), q_4(t), \lambda(t), \mu(t)\}$.

Notice that the first equation in \eqref{coneomsx} can be integrated to yield
\begin{equation}
\sigma'(t,x)=-\frac{1}{x^2}-\frac{1}{4x^2}\int_0^x ds\ \biggl\{s^2 \sigma(t,s)
\left((\phi'(t,s))^2+(\chi'(t,s))^2\right)
\biggr\}\,,
\eqlabel{delxsigma}
\end{equation}
where we used the asymptotes \eqref{basym}, implying that, $\sigma(t,x)$ is necessarily
positive,
\begin{equation}
\sigma'(t,x)<0 \,,
\eqlabel{delxsigma2}
\end{equation}
for all values of $x$. This fact will be important for the
dynamical entropy growth theorem,
see appendix \ref{enttheorem}.

The observables of
interest, \ie the energy density $\cale(t)$, the pressure $P(t)$, and the expectation
values of the operators $\calo_\phi(t)$ and $\calo_\chi(t)$ (dual to the bulk
scalars $\phi$ and $\chi$ correspondingly) can be computed following the
holographic renormalization of the model \cite{Buchel:2017lhu}:
\begin{equation}\begin{split}
\frac{2\kappa^2}{H^3}\ \cale(t)=& -4\mu(t) +\frac 13\ p_1^2+\left(\dd_1\ p_1^3+2\dd_2\ p_1 \right)\,,
\end{split}
\eqlabel{vev1}
\end{equation}
\begin{equation}\begin{split}
\frac{2\kappa^2}{H^3}\ P(t)=& -2\mu(t) -\frac 13 p_1^2+\frac 12 p_1 ( p_2(t)+\lambda(t) p_1)
+\left(-\dd_1\ p_1^3-2\dd_2\ p_1 \right)\,,
\end{split}
\eqlabel{vev2}
\end{equation}
\begin{equation}
\begin{split}
\frac{2\kappa^2}{H^2}\ \calo_\phi(t)=&-p_2(t)-\lambda(t) p_1 +p_1+\left(3 \dd_1\ p_1^2+6\dd_2 \right)\,,
\end{split}
\eqlabel{vev3}
\end{equation}
\begin{equation}
\begin{split}
\frac{2\kappa^2}{H^4}\ \calo_\chi(t)=&q_4(t)\,,
\end{split}
\eqlabel{vev4}
\end{equation}
where the terms in brackets, depending on arbitrary constants $\{\dd_1,\dd_2\}$, encode the renormalization scheme 
ambiguities. Independent of the renormalization scheme, 
these expectation values 
satisfy the expected conformal Ward identity
\begin{equation}
\begin{split}
&-\cale+2P=-p_1 H\ \calo_\phi\,.
\end{split}
\eqlabel{ward}
\end{equation}
Furthermore, the conservation of the stress-energy tensor 
\begin{equation}\eqlabel{cons}
\frac{d\cale}{dt}+2\ (\cale+P)=0\,,
\end{equation}
is a consequence of the momentum constraint \eqref{coneoms2}:   
\begin{equation}
\begin{split}
&0=\dot \mu+  3\mu-\frac {p_1}{4}\left(p_2+\lambda p_1\right)\,.
\end{split}
\eqlabel{ward2}
\end{equation}
From now on we choose a scheme with $\dd_i=0$.

The remaining  parameters $\{p_1, \lambda(t)\}$ have the following interpretation:
\begin{itemize}
\item $p_1$ is identified with the deformation mass scale $\Lambda$
\eqref{defformation},
\begin{equation}
p_1=\frac{\Lambda}{H}\,,
\eqlabel{phidat}
\end{equation}
\item   $\lambda(t)$ is the residual radial coordinate diffeomorphism parameter
\cite{Chesler:2013lia},
\begin{equation}
\frac rH \equiv\frac 1x\to \frac 1x+\lambda(t) \;,
\eqlabel{resdiffeo}
\end{equation}
which can be adjusted to keep the apparent horizon at a fixed location, which in our
case will be $r=r_{AH}=H$, equivalently, $x_{AH}=1$:
\begin{equation}
\biggl(\del_\tau +A(\tau,r)\ \del_r\ \biggr) \Sigma(t,r)\bigg|_{r_{AH}}\ \equiv\
 H^2 e^{t} \biggl(d_+\sigma(t,x)+\sigma(t,x)\biggr)\bigg|_{x_{AH}}=0 \;.
\eqlabel{ldata}
\end{equation}
\end{itemize}

To initialize evolution at $t=0$, we  provide the bulk scalar profiles,
\begin{equation}
\phi(t=0,x)=p_1 x+\calo\left(x^2\right)\,,\qquad
\chi(t=0,x)=\calo\left(x^4\right) \;,
\eqlabel{initphichi}
\end{equation} 
along with the values of $\{p_1,\mu(t=0)\}$, specifying the dual $QFT_3$ mass scale $\Lambda$ \eqref{phidat} and the initial state energy density  
$\cale(t=0)$ \eqref{vev1}.
The Hamiltonian constraint equation in \eqref{coneomsx} is then used to determine an initial profile $\sigma(t=0,x)$. 
Eqs.~\eqref{evolveomsx} are then employed to evolve such data \eqref{initphichi} in time. 
The second constraint in \eqref{coneomsx}, representing the conservation
of the boundary stress-energy tensor, is enforced requiring that 
a dynamical variable $\mu(t)$ in the asymptotic expansion of $a$, see \eqref{basym},
evolves following \eqref{ward2}.

Details of the numerical implementation, specific choices of the initial conditions \eqref{initphichi} used, 
and code convergence tests can be found in appendix \ref{num}.

\section{$\zet_2$-symmetric DFP and its fluctuations}\label{z2u}

\subsection{$\zet_2^\chi$-symmetric DFP}\label{dfpu}

$\zet_2^\chi$-symmetric dynamical fixed point is the $t\to \infty$ (late-time)
solution of \eqref{evolveomsx}-\eqref{coneomsx} with $\chi(t,x)\equiv 0$.
Introducing
\begin{equation}
\lim_{t\to \infty} \{\sigma,a,\phi\}(t,x)\ =\ \biggl\{\frac{\calf(x)}{x}\,,\,
\frac{\calg(x)}{2x^2}\,,\, p(x)\biggr\} \,,
\eqlabel{vdfp}
\end{equation}
we find from \eqref{evolveomsx}-\eqref{coneomsx}
\begin{equation}
\begin{split}
&0=\calf''+\frac14 \calf\ (p')^2\,,
\end{split}
\eqlabel{vu1}
\end{equation}
\begin{equation}
\begin{split}
&0=p''-\frac{2 \calf x^2}{K} (p')^3+\frac{\calf p}{K} (p')^2+\frac1K \biggl(
-\frac{24 x^2 (\calf')^2}{\calf}-\frac{2 (2 \calf' x-\calf) p^2}{x}+24 (x-1) \calf'
\\
&+\frac{12 \calf}{x}\biggr) p'-\frac{12 p}{\calf x^2 K} (\calf' x-\calf)^2\,,
\end{split}
\eqlabel{vu2}
\end{equation}
where
\begin{equation}
\begin{split}
&K=-\calf p^2-12 \calf' x^2+6 \calf (2 x-1)\,,
\end{split}
\eqlabel{defku}
\end{equation}
along with the algebraic equation for $\calg$:
\begin{equation}
\begin{split}
&\calg=\frac{2 \calf K}{\calf^2 x^2 (p')^2-12 (\calf' x-\calf)^2}\,.
\end{split}
\eqlabel{vu3}
\end{equation}
The late-time apparent horizon is located at $x_{AH}$, such that, see \eqref{ldata},
\begin{equation}
0=-\frac{\calg}{2} \frac{d\sigma}{dx}+ \sigma\bigg|_{x_{AH}}\,.
\eqlabel{uah}
\end{equation}
Since $\sigma'<0$ \eqref{delxsigma2}, it is clear from \eqref{uah} that
$\calg(x_{AH})<0$. However, as $\calg(x\to 0)=1$, see \eqref{basym},
this implies that $\calg$ vanishes at certain $x_s\in (0,x_{AH})$.
Vanishing of $\calg$ arises from vanishing of $K$ \eqref{defku}, $K(x_s)=0$,
making the bulk scalar equation \eqref{vu2} 'singular' at this point ---
note the factor $K$ in the denominator of some terms. This 'singularity' is
spurious, and solutions $\{\calf,p,\calg\}$ are in fact smooth for
$x\in[0,x_{AH}]$. To solve for a dynamical fixed point \eqref{vu1}-\eqref{vu3}, it
is convenient to use the residual diffeomorphism parameter $\lambda$ so that 
$x_s$, rather than $x_{AH}$, is kept fixed (as one varies $p_1$):
\begin{equation}
x_s=\frac 13\,.
\eqlabel{xsdef}
\end{equation}
Thus, we seek solutions of \eqref{vu1}-\eqref{vu3} subject to the
following boundary conditions:
\nxt in the UV\footnote{Compare with \eqref{basym}.}, \ie as $x\to 0_+$, 
\begin{equation}
\begin{split}
&\calf=1+\lambda\ x-\frac18 p_1^2\ x^2+\left(-\frac{1}{24} \lambda p_1^2-\frac16
p_1 p_2
\right)\ x^3+\calo(x^4)\,,
\\
&p=p_1\ x+p_2\ x^2+\left(
\frac14 p_1^3-2 \lambda p_2+2 p_2+2 \lambda p_1-\lambda^2 p_1
\right)\ x^3+\calo(x^4)\,,
\\
&\calg=1+(-2+2 \lambda)\ x+\left(
-\frac14 p_1^2+\lambda^2-2 \lambda\right)\ x^2
+\left(
-\frac13 \lambda p_1^2+\frac13 p_1^2-\frac13 p_1 p_2
\right)\ x^3
\\&+\left(-\frac{1}{24} p_1^4-\frac16 p_2^2-\frac13 \lambda p_1^2
+\frac16 \lambda^2 p_1^2\right)\ x^4+\calo(x^5)\,;
\end{split}
\eqlabel{uvu}
\end{equation}
\nxt in the IR, \ie,  as $y\equiv \frac 13-x\to 0_+$, 
\begin{equation}
\begin{split}
&\calf=f_0^s+\frac34 f_0^s ((p^s_0)^2+2)\ y+\calo(y^2)\,,
\\
&p=p^s_0-\frac92 p^s_0\ y+\calo(y^2)\,,
\\
&\calg=2 y+\left(3+\frac32 (p^s_0)^2\right)\ y^2+\calo(y^3)\,.
\end{split}
\eqlabel{sinu}
\end{equation}
For a fixed $p_1$, alternatively the mass scale $\Lambda$ \eqref{phidat},
the DFP gravitational dual is characterized by $\{p_2,\lambda,f^s_0,p^s_0\}$ ---
precisely the number of parameters needed to solve the  pair of
the second order ODEs \eqref{vu1} and \eqref{vu2}.
Notice that $\mu$ is not an independent parameter --- from \eqref{ward2}
at the DFP $\dot\mu=0$ leading to
\begin{equation}
\mu=\frac{p_1}{12}(p_2+\lambda p_1)\,.
\eqlabel{mudfp}
\end{equation}

The solutions for $x\in[0,\frac 13)$ are then extended for
$x\in(\frac13,x_{AH}]$, solving \eqref{vu1} and \eqref{vu2}, with initial conditions
(see \eqref{sinu})
\begin{equation}
\begin{split}
&\calf=f_0^s+\frac34 f_0^s ((p^s_0)^2+2)\ \left(\frac 13-x\right)
+\calo\left(\left(\frac 13-x\right)^2\right)\,,
\\
&p=p^s_0-\frac92 p^s_0\ \left(\frac 13-x\right)+\calo\left(\left(\frac 13-x\right)^2\right)\,,
\\
&\calg=2 \left(\frac 13-x\right)+\left(3+\frac32 (p^s_0)^2\right)\ \left(\frac 13-x\right)^2+\calo\left(\left(\frac 13-x\right)^3\right)\,,
\end{split}
\eqlabel{sinu1}
\end{equation}
while monitoring the AH locator \eqref{uah}.

For a finite $p_1$, the data sets $\{p_2,\lambda,f^s_0,p^s_0\}$ are found numerically, using the shooting method developed in \cite{Aharony:2007vg}.
As originally presented in \cite{Buchel:2017lhu}, equations \eqref{vu1} and \eqref{vu2}
can be solved perturbatively in $p_1$:
\begin{equation}
\begin{split}
&\calf=1-x+ \frac{x (4 x-1)}{24(x-1)}\ p_1^2
+ \frac{x (4 x-1) (23 x^2-5 x-5)}{3456(x-1)^3}\ p_1^4
+  \frac{x}{6220800(x-1)^5} (49618 x^5\\
&-46133 x^4+9055 x^3-2745 x^2+3225 x-645)
\ p_1^6+\frac{x}{87787929600(x-1)^7}
\\&\times (239535208 x^7-287767231 x^6+66948945 x^5+34436913 x^4
-23724575 x^3\\
&+13512065 x^2-5761175 x+823025)\  p_1^8+\calo(p_1^{10})\,,
\end{split}
\eqlabel{fpert}
\end{equation}
\begin{equation}
\begin{split}
&p= \frac{x}{(1-x)}\ p_1-\frac{x^2 (2x-1)}{9(x-1)^3}\ p_1^3
- \frac{x^2 (875 x^3-647 x^2+9 x+51)}{12960(x-1)^5}\ p_1^5
\\&-\frac{x^2(822367 x^5-874077 x^4+196890 x^3+19590 x^2+17295 x-9265)}{32659200(x-1)^7} \
p_1^7+\calo(p_1^{9})\,,
\end{split}
\eqlabel{ppert}
\end{equation}
\begin{equation}
\begin{split}
&\calg=(1-3 x)\biggl(
1-x+\frac{(3 x-1) x}{12(x-1)}\ p_1^2
+ \frac{(3 x-1) (19 x^2-2 x-5) x}{1728(x-1)^3}\ p_1^4
+ \frac{x(3 x-1)}{622080(x-1)^5}\\
&\times(1937 x^4-1196 x^3+54 x^2-204 x+129)\  p_1^6
+\frac{(3 x-1)x}{6270566400(x-1)^7} (6281809 x^6\\
&-5336034 x^5-314745 x^4+1210180 x^3-631665 x^2+420030 x-117575)\  p_1^8
\biggr)+\calo(p_1^{10})\,,
\end{split}
\eqlabel{gpert}
\end{equation}
resulting in 
\begin{equation}
\begin{split}
&p_2=p_1-\frac19 p_1^3+\frac{17}{4320} p_1^5-\frac{1853}{6531840} p_1^7+\calo(p_1^9)\,,\\
&\lambda=-1+\frac{1}{24} p_1^2-\frac{5}{3456} p_1^4+\frac{43}{414720} p_1^6
-\frac{4703}{501645312} p_1^8
+\calo(p_1^{10})\,.
\end{split}
\eqlabel{p2l}
\end{equation}
The choice of $x_s$ in \eqref{xsdef} is motivated so that $\lambda(p_1=0)=-1$.
Given \eqref{p2l}, we can compute perturbatively in $p_1$ one-point correlation
functions of the stress energy-tensor and $\calo_\phi$, \eqref{vev1}-\eqref{vev3}:
\begin{equation}
\begin{split}
&\frac 1c\ \frac{\cale}{H^3}=\frac{1}{1152} p_1^2+\frac{5}{82944} p_1^4
-\frac{43}{19906560} p_1^6+\frac{4703}{30098718720} p_1^8+\calo(p_1^{10})\,,\\
&\frac 1c\ \frac{P}{H^3}=-\frac{1}{1152} p_1^2-\frac{5}{82944} p_1^4+\frac{43}{
19906560} p_1^6-\frac{4703}{30098718720} p_1^8+\calo(p_1^{10})\,,\\
&\frac 1c \frac{\calo_\phi}{H^2}=\frac{5}{27648} p_1^3-\frac{43}{6635520} p_1^5
+\frac{4703}{10032906240}
p_1^7+\calo(p_1^{9})\,,
\end{split}
\eqlabel{pertvevs}
\end{equation}
where $c$ is the central charge \eqref{c}.

The obtained data
sets are used to generate figs. \ref{epunbroken} and \ref{osunbroken}.

\subsection{Fluctuations about $\zet_2^\chi$-symmetric DFP}\label{dfpflu}

The DFP of section \ref{dfpu} has an unbroken $\zet_2^\chi$ symmetry.
The fluctuations about it decouple into sets, characterized by this
symmetry:
\nxt Fluctuations $\{\delta\cale,\delta P, \delta\calo_\phi\}$ preserve
$\zet_2^\chi$ symmetry, and are stable. They have been extensively studied in
\cite{Buchel:2017lhu}. 
\nxt Fluctuations $\delta \calo_\chi$ spontaneously break $\zet_2^\chi$ symmetry.
They end up being unstable, provided $\frac\Lambda H$ is sufficiently
large. 

The spectrum of $\zet_2^\chi$ symmetry breaking fluctuations
$\delta \calo_\chi$ is represented by the spectrum of the
quasinormal modes of the gravitational scalar $\chi$ in the
holographic DFP background \eqref{vu1}-\eqref{vu3}.
Introducing
\begin{equation}
\chi(x,t)=H_0(x)\ e^{-i \hw t}\,,\qquad \hw=\frac\omega H\,,
\eqlabel{flu1}
\end{equation}
we obtain from the linearization of the third equation in \eqref{evolveomsx}
about \eqref{vu1}-\eqref{vu3}:
\begin{equation}
0=H_0''+\biggl(\frac{2 \calf'}{\calf}+\frac{\calg'}{\calg}+\frac{2 i \hw}{\calg}
-\frac2x-\frac{2}{\calg}\biggr) H_0'+2 \biggl(
\frac{i \calf' \hw}{\calg \calf}-\frac{p^2 g}{\calg x^2}
-\frac{i \hw}{\calg x}-\frac{2}{\calg x^2}
\biggr) H_0\,.
\eqlabel{flu2}
\end{equation}
The spectrum is determined solving for $H_0$, subject to the following
asymptotics (without the loss of generality we normalized the linearized
fluctuations so that $H_{0,4}=1$, see \eqref{vev4}):
\nxt in the UV, \ie as $x\to 0_+$,
\begin{equation}
H_0=x^4\ \biggl(\underbrace{H_{0,4}}_{=1}+(4-4 \lambda-i \hw)\ x
+\calo(x^2)\biggr)\,;
\eqlabel{flu3}
\end{equation}
\nxt in the IR, \ie as $y\equiv \frac 13-x\to 0_+$,
\begin{equation}
H_0=\left(\frac 13-y\right)^4\ \biggl(H^s_{0,0}-\frac{3H^s_{0,0} (i \hw (p^s_0)^2
+12 (p^s_0)^2 g-(10 i) \hw+56)}{4(i \hw-2)}\ y+\calo(y^2)\biggr)\,.
\eqlabel{flu4}
\end{equation}
Note that there are two parameters characterizing a QNM: $\{\hw, H^s_{0,0}\}$ ---
precisely as needed to solve a single second order ODE \eqref{flu2}.

For a finite $p_1$, the spectrum is computed numerically, see
fig.~\ref{wunbroken}. We can also compute the spectrum perturbatively
in $p_1$, using \eqref{fpert}-\eqref{gpert},
\begin{equation}
\begin{split}
&\hw_n=-i n+\calo(p_1^2)\,,\qquad n\in \mathbb{N} \ge 4\,,
\end{split}
\eqlabel{pertwu}
\end{equation}
\eg
\begin{equation}
\begin{split}
&\hw_4=-i \biggl(
4+\left(\frac14 g-\frac{1}{60}\right) p_1^2+\left(
\frac{3}{160} g^2-\frac{157}{5400} g+\frac{23}{21600}
\right) p_1^4+\biggl(-\frac{9073}{108864000}
\\&+\frac{17}{4800} g^3+\frac{388879}{108864000} g
-\frac{4217}{648000} g^2\biggr) p_1^6+\biggl(
\frac{7693913}{1097349120000}-\frac{4356421}{2177280000} g^3
\\&-\frac{1815343}{4064256000} g+\frac{1621804901}{1097349120000} g^2
+\frac{2053}{2304000} g^4\biggr) p_1^8+\calo(p_1^{10})
\biggr)\,,
\end{split}
\eqlabel{pertwu4}
\end{equation}
with $n>4$ perturbative spectra readily obtained as well.

\section{DFP with spontaneously broken $\zet_2$ symmetry
and its fluctuations}\label{z2b}

\subsection{$\zet_2^\chi$-broken DFP}\label{dfpb}

A dynamical fixed point with spontaneously broken $\zet_2^\chi$
symmetry is the $t\to \infty$ (late-time)
solution of \eqref{evolveomsx}-\eqref{coneomsx} with $\chi(t,x)\ne 0$.
Introducing
\begin{equation}
\lim_{t\to \infty} \{\sigma,a,\phi,\chi\}(t,x)\ =\ \biggl\{\frac{\calf(x)}{x}\,,\,
\frac{\calg(x)}{2x^2}\,,\, p(x)\,,\, q(x)\biggr\} \,,
\eqlabel{bdfp}
\end{equation}
we find  from \eqref{evolveomsx}-\eqref{coneomsx}
\begin{equation}
\begin{split}
&0=\calf''+\frac14 \calf \left((p')^2+(q')^2\right)\,,
\end{split}
\eqlabel{vb1}
\end{equation}
\begin{equation}
\begin{split}
&0=p''-\frac{2 \calf x^2}{K} (p')^3-\frac{\calf p (q^2 g-1)}{K} (p')^2+\frac1K \biggl(
-2 \calf x^2 (q')^2-\frac{24 x^2 (\calf')^2}{\calf} \\&
+\frac{2 (2 \calf' x-\calf)}{x}
(p^2 q^2 g-p^2+2 q^2)+24 (x-1) \calf'+\frac{12 \calf}{x}
\biggr) p'\\
&+\frac{p (-\calf^2 (q')^2 x^2+12 (\calf' x-\calf)^2) (q^2 g-1)}{\calf x^2 K}\,,
\end{split}
\eqlabel{vb2}
\end{equation}
\begin{equation}
\begin{split}
&0=q''-\frac{2 \calf x^2}{K} (q')^3-\frac{\calf q (p^2 g+2)}{K} (q')^2+\frac1K \biggl(
-2 \calf x^2 (p')^2-\frac{24 x^2(\calf')^2}{\calf} \\
&+\frac{2 (2 \calf' x-\calf)}{x}
(p^2 q^2 g-p^2+2 q^2)+24 (x-1) \calf'+\frac{12 \calf}{x}\biggr) q'
\\&+\frac{q (-\calf^2 x^2 (p')^2+12 (\calf' x-\calf)^2) (p^2 g+2)}
{\calf x^2 K}\,,
\end{split}
\eqlabel{vb3}
\end{equation}
where
\begin{equation}
\begin{split}
&K=\calf (p^2 q^2 g-p^2+2 q^2)-12 \calf' x^2+6 \calf (2 x-1)\,,
\end{split}
\eqlabel{defkb}
\end{equation}
along with the algebraic equation for $\calg$:
\begin{equation}
\begin{split}
&\calg=\frac{2 \calf K}{\calf^2 x^2 ((p')^2+(q')^2)-12 (\calf' x-\calf)^2}\,.
\end{split}
\eqlabel{vb4}
\end{equation}
The location of the apparent horizon $x_{AH}$ is given by the same expression
as \eqref{uah}. As for the $\zet_2^{\chi}$-symmetric DFP
in section \ref{z2u}, the vanishing of $K$ (correspondingly $\calg$)
at $x_s\in (0,x_{AH})$ renderes the equations for the bulk scalars
\eqref{vb2} and \eqref{vb3} singular at $x_s$. We use the residual
diffeomorphism parameter $\lambda$, see \eqref{resdiffeo}, to keep
the location of $x_s$ fixed as in \eqref{xsdef}, as one varies $p_1$. 

We seek solutions of \eqref{vb1}-\eqref{vb4} subject to the
following boundary conditions:
\nxt in the UV, \ie as $x\to 0_+$, 
\begin{equation}
\begin{split}
&\calf=1+\lambda\ x-\frac18 p_1^2\ x^2+\left(-\frac{1}{24} \lambda p_1^2-\frac16
p_1 p_2
\right)\ x^3+\calo(x^4)\,,
\\
&p=p_1\ x+p_2\ x^2+\left(
\frac14 p_1^3-2 \lambda p_2+2 p_2+2 \lambda p_1-\lambda^2 p_1
\right)\ x^3+\calo(x^4)\,,
\\
&q=q_4\ x^4+4 q_4(1- \lambda)\ x^5
+q_4 \left(
\frac{p_1^2 g}{7}+10 \lambda^2 +\frac{5p_1^2}{14}-20 \lambda +\frac{80}{7} 
\right)\ x^6+\calo(x^7)\,,
\\
&\calg=1+(-2+2 \lambda)\ x+\left(
-\frac14 p_1^2+\lambda^2-2 \lambda\right)\ x^2
+\left(
-\frac13 \lambda p_1^2+\frac13 p_1^2-\frac13 p_1 p_2
\right)\ x^3
\\&+\left(-\frac{1}{24} p_1^4-\frac16 p_2^2-\frac13 \lambda p_1^2
+\frac16 \lambda^2 p_1^2\right)\ x^4+\calo(x^5)\,;
\end{split}
\eqlabel{uvb}
\end{equation}
\nxt in the IR, \ie,  as $y\equiv \frac 13-x\to 0_+$, 
\begin{equation}
\begin{split}
&\calf=f^s_0-\frac34 f^s_0 \left(
(p^s_0)^2 (q^s_0)^2 g-(p^s_0)^2+2 (q^s_0)^2-2\right)\ y+\calo(y^2)\,,
\\
&p=p^s_0+\left(\frac92 p^s_0 (q^s_0)^2 g-\frac92 p^s_0\right)\ y+\calo(y^2)\,,
\\
&q=q^s_0+\left(\frac92 (p^s_0)^2 q^s_0 g+9 q^s_0\right)\ y+\calo(y^2)\,,
\\
&\calg=2 y+\left(\frac32 (p^s_0)^2(1-(q^s_0)^2 g)-3 (q^s_0)^2+3\right)\ y^2+\calo(y^3)\,.
\end{split}
\eqlabel{sinb}
\end{equation}
For a fixed $p_1$, alternatively the mass scale $\Lambda$ \eqref{phidat},
the DFP gravitational dual is characterized by $\{p_2,q_4,
\lambda,f^s_0,p^s_0,q^s_0\}$ ---
precisely the number of parameters needed to solve three 
second order ODEs \eqref{vb1}-\eqref{vb3}.
Notice that $\mu$ is not an independent parameter --- from \eqref{ward2}
at the DFP $\dot\mu=0$ leading to \eqref{mudfp}.

The solutions for $x\in[0,\frac 13)$ are then extended for
$x\in(\frac13,x_{AH}]$, solving \eqref{vb1}-\eqref{vb3}, with initial conditions
(see \eqref{sinb})
\begin{equation}
\begin{split}
&\calf=f^s_0-\frac34 f^s_0 \left(
(p^s_0)^2 (q^s_0)^2 g-(p^s_0)^2+2 (q^s_0)^2-2\right)\ \left(\frac 13-x\right)
+\calo\left(\left(\frac 13-x\right)^2\right)\,,
\\
&p=p^s_0+\left(\frac92 p^s_0 (q^s_0)^2 g-\frac92 p^s_0\right)\ \left(\frac 13-x\right)+\calo\left(\left(\frac 13-x\right)^2\right)\,,
\\
&q=q^s_0+\left(\frac92 (p^s_0)^2 q^s_0 g+9 q^s_0\right)\ \left(\frac 13-x\right)+\calo\left(\left(\frac 13-x\right)^2\right)\,,
\\
&\calg=2 \left(\frac 13-x\right)+\left(
\frac32 (p^s_0)^2(1-(q^s_0)^2 g)-3 (q^s_0)^2+3\right)\ \left(\frac 13-x\right)^2+\calo\left(\left(\frac 13-x\right)^3\right)\,,
\end{split}
\eqlabel{sinb1}
\end{equation}
while monitoring the AH locator \eqref{uah}.

The data sets $\{p_1,p_2,q_4,\lambda,f^s_0,p^s_0,q^s_0\}$ are found numerically:
$\zet_2^\chi$-broken DFPs do not extend to $p_1\to 0$, making
analytic perturbative analysis impossible. The obtained data
sets are used to generate figs. \ref{epbroken} and \ref{sbroken}.

\subsection{Fluctuations about $\zet_2^\chi$-broken DFP}\label{dfpbfl}

Introducing
\begin{equation}
\begin{split}
&\chi=q(x)+ H_0(x)\ e^{-i\hw t}\,,\qquad \phi(t,x)=p(x)+ H_1(x)\ e^{-i\hw t}\,,\\
&\sigma(t,x)=\frac1x\left(\calf(x)+ H_2(x)\ e^{-i\hw t}\right)\,,\qquad 
a(t,x)=\frac{1}{2x^2}\left(\calg(x)+ H_3(x)\ e^{-i\hw t} \right)\,,
\end{split}
\eqlabel{defflb}
\end{equation}
we obtain from the linearization of \eqref{evolveomsx}-\eqref{coneomsx}
with respect to $ H_i$,
\begin{equation}
\begin{split}
&0=H_0''+\left(
\frac{2 i \hw}{\calg}+\frac{2 \calf'}{\calf}+\frac{\calg'}{\calg}-\frac2x-\frac2\calg
\right)\ H_0'
+\frac{2 q'}{\calf}\ H_2'+\frac{q'}{\calg}\ H_3'
+\biggl(
\left(\frac{2 i \calf'}{\calg \calf}-\frac{2 i}{\calg x}\right) \hw
\\&-\frac{2 p^2 g}{\calg x^2}-\frac{4}{\calg x^2}
\biggr)\ H_0
-\frac{4 p q g}{\calg x^2}\ H_1+\biggl(
\frac{2 i q' \hw}{\calg \calf}-\frac{2 q' \calf}{\calf^2}
\biggr)\ H_2
+\biggl(
\frac{2 p^2 q g}{\calg^2 x^2}-\frac{q' \calg'}{\calg^2}
+\frac{2 q'}{\calg^2}\\&+\frac{4 q}{\calg^2 x^2}
\biggr)\ H_3\,,
\end{split}
\eqlabel{flh0}
\end{equation}
\begin{equation}
\begin{split}
&0=H_1''+\biggl(
\frac{2 i \hw}{\calg}+\frac{2 \calf'}{\calf}+\frac{\calg'}{\calg}-\frac2x
-\frac2\calg\biggr)\
H_1'
+\frac{2 p'}{\calf}\ H_2'
+\frac{p'}{\calg}\ H_3'-\frac{4 p q g}{\calg x^2}\ H_0
\\&+\biggl(
\left( \frac{2 i \calf'}{\calg \calf}-\frac{2 i}{\calg x}\right) \hw
-\frac{2 q^2 g}{\calg x^2}+\frac{2}{\calg x^2}
\biggr)\ H_1
+\biggl(
\frac{2 i p' \hw}{\calg \calf}-\frac{2 p' \calf'}{\calf^2}\biggr)\ H_2
+\biggl(
\frac{2 p q^2 g}{\calg^2 x^2}-\frac{p' \calg'}{\calg^2}
\\&+\frac{2 p'}{\calg^2}-\frac{2 p}{\calg^2 x^2}\biggr)\ H_3\,,
\end{split}
\eqlabel{flh1}
\end{equation}
\begin{equation}
\begin{split}
&0=H_2''+\frac12 \calf q'\ H_0'+\frac12 \calf p'\ H_1'+\frac14
\left( (p')^2+(q')^2 \right)\ H_2\,,
\end{split}
\eqlabel{flh2}
\end{equation}
where we kept $H_3$ dependence to avoid cluttering the formulas,
although $H_3'$ and $H_3$ can be expressed algebraically through  
$\{H_0,H_1,H_2\}$ and their first derivatives:
\begin{equation}
\begin{split}
&H_3'=\frac{ x \calf \calg q' (\calg-1)}{2M}\ H_0'
+\frac{ x \calf \calg p' (\calg-1)}{2M}\ H_1'
+\frac1M \biggl(
-\frac{2 \calg x (\calg-1) \calf'}{\calf}+x (1-\calg) \calg'
\\&+2 i \hw x +4 \calg^2+2 x \calg-4 \calg-4 x\biggr)\ H_2'
+\frac1M \biggl(
\frac12 i \calg \calf \hw x q'-\frac{q \calf (p^2 g+2) (\calg-1)}{x}\biggr)\ H_0\\
&+\frac1M \biggl(
\frac12 i \calg \calf \hw x p'-\frac{p \calf (q^2 g-1) (\calg-1)}{x}\biggr)\ H_1
+\frac1M \biggl(
\frac{\calg x (\calg-1)(\calf')^2}{\calf^2}\\
&+\frac14 \calg x (\calg-1) ((p')^2+(q')^2)
-\frac{2 i \hw x (\calg-1) \calf'}{\calf}+(-i \hw x+\calg+x-1) \calg'
-2 x \hw^2\\
&+4 i (\calg-x-1) \hw-\frac{3 \calg^2}{x}-\frac{\calg(p^2 (q^2 g-1)+2 q^2+8 x-12)}{2x} 
\\&+\frac{(p^2 (q^2 g-1)+2 q^2+4 x^2+8 x-6)}{2x}
\biggr)\ H_2
+\frac1M \biggl(
\frac14 \calf x (2 \calg-1) ((p')^2+(q')^2)\\
&-\frac{x (2 \calg-1)(\calf')^2}{\calf} +(i \hw x-\calg' x+8 \calg+2 x-4) \calf'
+(\calf-x \calf') \calg'+x \calg' \calf'-i \hw \calf
\\&-\frac{\calf (p^2 (q^2 g-1)+2 q^2+12 \calg+8 x-12)}{2x}
\biggr)\ H_3\,,
\end{split}
\eqlabel{flh3}
\end{equation}
\begin{equation}
\begin{split}
&H_3=\frac1N \biggl(
-2 \calg q' \calf^3 x^3\ H_0'-2 p' \calg \calf^3 x^3\ H_1'
+(8 \calf \calg (2-i \hw) x^3 \calf'+4 \calf^2 (\calg' x-6 \calg
\\&-4 x
+2 i \hw (\calg+x)) x^2)\ H_2'
-2 x \calf^2 (i \calg q' (x \calf'-\calf) \hw x-2 q \calf (p^2 g+2)) \ H_0
\\&-2 x \calf^2 (i p' \calg \hw x (x \calf'-\calf)-2 p (q^2 g-1) \calf)\  H_1
+(
-\calf^2 \calg x^3 ((p')^2+(q')^2)-4 \calg x^3 (\calf')^2
\\&
+4 \calf x^3(i \calg' \hw+6 i \hw+2 \hw^2-\calg'-2) \calf'
-4 i \calf^2 x^2 \hw \calg'-2 x \calf^2
(p^2 (1-q^2 g)+16 i \hw x\\
&+4 x \hw^2-2 q^2-6 \calg-12 x+6)
)\ H_2
\biggr)\,,
\end{split}
\eqlabel{flh4}
\end{equation}
with
\begin{equation}
\begin{split}
&M=x \calf' (\calg-1)-\calf (\calg+x-1)\,,
\end{split}
\eqlabel{defm}
\end{equation}
and
\begin{equation}
\begin{split}
&N=-4 \calg x^3 (\calf')^3
+4  \calf x^2 (i \hw x-\calg' x+5 \calg+x) (\calf')^2
-2 \calf^2 x (p^2 q^2 g+4 i \hw x-p^2+2 q^2\\
&-4 \calg' x+14 \calg+4 x-6) \calf'
+\calf^2 x^2 (x \calf' \calg-\calf \calg+x \calf)
((p')^2+(q')^2)\\
&+2 \calf^3 (p^2 (q^2 g-1)+2 i \hw x+2 q^2-2 \calg' x+6 \calg+2 x-6)\,.
\end{split}
\eqlabel{defn}
\end{equation}
We explicitly verified that \eqref{flh3} and \eqref{flh4} are consistent,
given \eqref{flh0}-\eqref{flh2} and the  background equations
\eqref{vb1}-\eqref{vb4}. 

The spectrum is determined solving for $H_i$,
(without the loss of generality we normalized the linearized
fluctuations so that $H_{0,4}=1$, see \eqref{vev4}):
\nxt in the UV, \ie as $x\to 0_+$,
\begin{equation}
H_0=x^4\ \biggl(\underbrace{H_{0,4}}_{=1}+(4-4 H_{2,1} q_4-i \hw-4 \lambda)\ x
+\calo(x^2)\biggr)\,,
\eqlabel{flb1}
\end{equation}
\begin{equation}
H_1=H_{1,2}\ x^2+((2p_1-i \hw  p_1-2  \lambda p_1-2p_2) H_{2,1}+(2-i \hw-2\lambda) H_{1,2}
)\ x^3
+\calo(x^4)\,,
\eqlabel{flb2}
\end{equation}
\begin{equation}
H_2=H_{2,1}\ x+\left(-\frac{p_1}{6} H_{1,2}-\frac{p_1^2}{24} H_{2,1} \right)\ x^3
+\calo(x^4)\,;
\eqlabel{flb3}
\end{equation}
\nxt in the IR, \ie as $y\equiv \frac 13-x\to 0_+$,
\begin{equation}
\begin{split}
&H_0=H_{0,0}^s+\frac{3}{4 f_0^s (i \hw-2)} \biggl(
\biggl(i f_0^s ((p_0^s)^2 ((q_0^s)^2 g-1)+2 (q_0^s)^2-6) \hw-12 f_0^s
((p_0^s)^2 g\\
&+2)\biggr)\ H_{0,0}^s-24 f_0^s p_0^s q_0^s g\ H_{1,0}^s-\biggl(6 q_0^s ((p_0^s)^2 g+2) \hw^2+12 i q_0^s
((p_0^s)^2 g+2) \hw\biggr)\ H_{2,0}^s
\biggr)\ y\\
&+\calo(y^2)\,,
\end{split}
\eqlabel{flb4}
\end{equation}
\begin{equation}
\begin{split}
&H_1=H_{1,0}^s+\frac{3}{4 f_0^s (i \hw-2)} \biggl(
-24 f_0^s p_0^s q_0^s g\ H_{0,0}^s+\biggl(
i f_0^s ((p_0^s)^2 (q_0^s)^2 g
-(p_0^s)^2+2 (q_0^s)^2-6) \hw\\
&-12 f_0^s ((q_0^s)^2 g-1)\biggr)\ H_{1,0}^s
-\biggl(6 p_0^s ((q_0^s)^2 g-1) \hw^2
+12 i p_0^s ((q_0^s)^2 g-1) \hw\biggr)\ H_{2,0}^s
\biggr)\ y
\\
&+\calo(y^2)\,,
\end{split}
\eqlabel{flb5}
\end{equation}
\begin{equation}
\begin{split}
&H_2=H_{2,0}^s+\frac{3}{4 f_0^s (i \hw-3)} \biggl(
6 (f_0^s)^2 q_0^s ((p_0^s)^2 g+2)\ H_{0,0}^s+(6 p_0^s ((q_0^s)^2 g-1)
(f_0^s)^2)\ H_{1,0}^s\\&+\biggl(
f_0^s \hw^2 (((q_0^s)^2 g-1) (p_0^s)^2+2 (q_0^s)^2-6)+(2 i) f_0^s
(((q_0^s)^2 g-1) (p_0^s)^2+2 (q_0^s)^2-8) \hw\\&+3 f_0^s (((q_0^s)^2 g-1)
(p_0^s)^2+2 (q_0^s)^2-2)\biggr)\ H_{2,0}^s
\biggr)\ y
+\calo(y^2)\,.
\end{split}
\eqlabel{flb6}
\end{equation}
Note that there are 6 parameters characterizing a QNM:
\begin{equation}
\{\hw\,,\, H_{1,2}\,,\, H_{2,1}\,,\,  H^s_{0,0}\,,\,  H^s_{1,0}\,,\,  H^s_{2,0}\}\,,
\eqlabel{parb}
\end{equation}
precisely as needed to solve three second order ODE \eqref{flh0}-\eqref{flh2}.

The unstable QNM is computed numerically, and is presented in
fig.~\ref{wbroken}.

\section{Dynamical entropy  and the entanglement entropy
of a DFP}\label{enttheorem}

\subsection{Dynamical entropy from the holographic dual}\label{theo1}

One of the advantages of the holographic formulation of a QFT dynamics is the natural definition 
of its far-from-equilibrium entropy density. A gravitational geometry \eqref{EFmetric} 
has an apparent horizon located at $r=r_{AH}$, where \cite{Chesler:2013lia}
\begin{equation}
d_+\Sigma\bigg|_{r=r_{AH}}=0\,.
\eqlabel{defhorloc}
\end{equation} 
Following \cite{Booth:2005qc,Figueras:2009iu} we associate the non-equilibrium  entropy density $s$
of the boundary QFT  with the Bekenstein-Hawking entropy density of the apparent horizon  
\begin{equation}
e^{2H\tau} s =\frac {2\pi}{\kappa^2}\ {\Sigma^2}\bigg|_{r=r_{AH}}\,.
\eqlabel{as}
\end{equation}
Using the holographic background equations of motion
\eqref{evolveoms}-\eqref{coneoms2} 
we find 
\begin{equation}
\frac{d(e^{2H\tau} s)}{d\tau}=\frac{2\pi}{\kappa^2}\ (\Sigma^2)'\ \frac{
 (d_+\phi)^2+(d_+\chi)^2}{\phi^2+6-2 \chi^2-g\phi^2 \chi^2}\bigg|_{r=r_{AH}}\,.
\eqlabel{dasdt}
\end{equation}
Following \cite{Buchel:2017pto} it is easy to  prove that the
comoving entropy production rate as defined by \eqref{dasdt}
is non-negative, \ie 
\begin{equation}
\frac{d(e^{2H\tau} s)}{d\tau}\ge 0\,,
\eqlabel{dasdt2}
\end{equation}
in holographic dynamics governed by \eqref{evolveoms}-\eqref{coneoms2}:
\nxt In \eqref{delxsigma2} we showed that $\del_x\sigma<0$,
thus
\begin{equation}
\del_r\Sigma>0\,,
\eqlabel{dpsigma0}
\end{equation}
for all times. 
\nxt Apparent horizon is defined as the innermost (with respect to the boundary)
coordinate location $r=r_{AH}$, where $d_+\Sigma(\tau,r)$ vanishes.
Notice from \eqref{basym},
\begin{equation}
d_+\Sigma(\tau,r)=\frac{e^{H\tau}r^2}{2}+\calo(r)\ >\ 0\,,\qquad r\to \infty\,.
\eqlabel{asymdpsigma}
\end{equation}
Thus, assuming the analyticity of the background geometry,
\begin{equation}
d_+\Sigma(\tau,r)>0\,,\qquad r>r_{AH}\qquad \Longrightarrow\qquad
\left(d_+\Sigma\right)'(\tau,r)\bigg|_{r=r_{AH}}\ge 0\,.
\eqlabel{dpsigma2}
\end{equation}
The first evolution equation in \eqref{evolveoms}, evaluated
at the apparent horizon, \ie when $d_+\Sigma=0$, reads
\begin{equation}
\begin{split}
&0=\left(d_+\Sigma\right)'-\frac\Sigma4(\phi^2+6-2\chi^2-g\phi^2\chi^2)\bigg|_{r=r_{AH}+0}\qquad \Longrightarrow\\
&(\phi^2+6-2\chi^2-g\phi^2\chi^2)\bigg|_{r=r_{AH}+0}\ge 0 \,,
\end{split}
\eqlabel{dpsigma3}
\end{equation}
given \eqref{dpsigma2}.
\nxt Combining \eqref{dpsigma0} and \eqref{dpsigma3}, we arrive from
\eqref{dasdt} to \eqref{dasdt2}.

\subsection{Entanglement entropy density of a DFP}\label{eed}

\begin{figure}[h]
\begin{center}
\psfrag{p}[cc][][1][0]{$\frac{\Lambda}{H}$}
\psfrag{d}[bb][][1][0]{$\delta_{ent}$}
\psfrag{r}[tt][][1][0]{$\delta_{ent}$}
\includegraphics[width=3in]{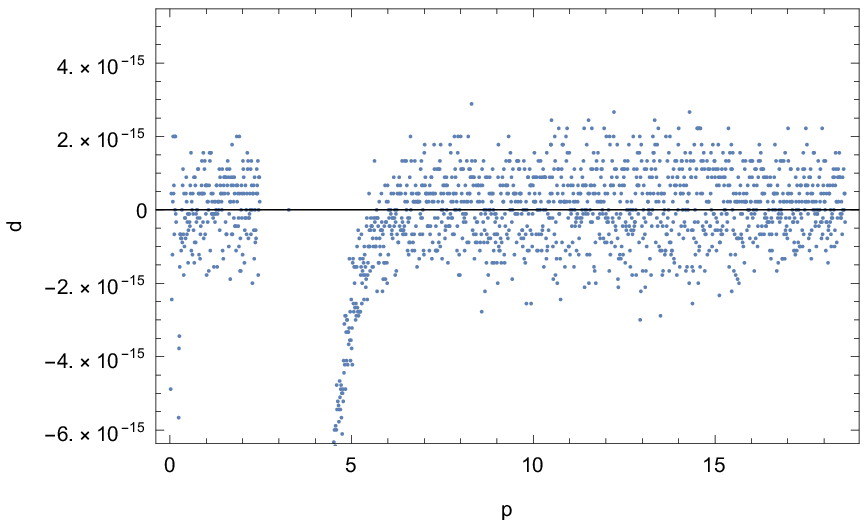}
\includegraphics[width=3in]{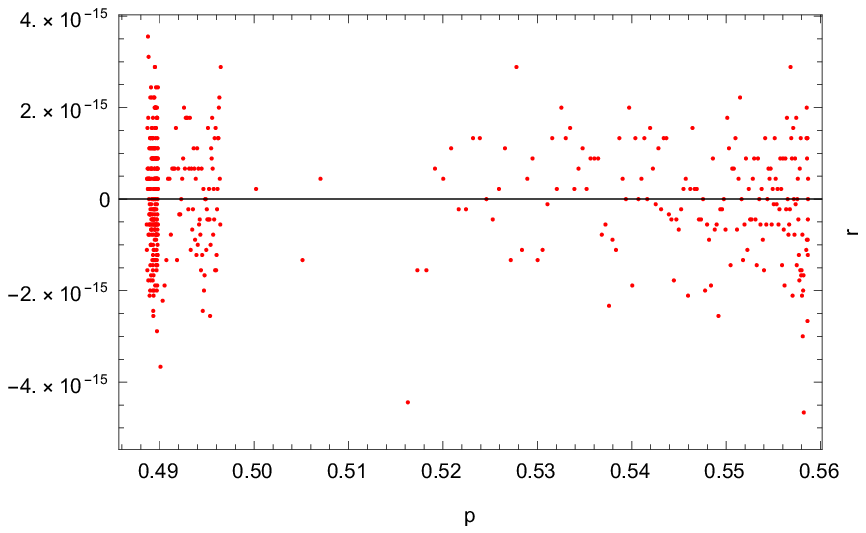}
\end{center}
  \caption{Comparison of the vacuum entanglement entropy density of the \dfps
(the left panel) and the \dfpb (the right panel) computed from \eqref{defsent}
and \eqref{defsent2}. See \eqref{errsent} for the definition of $\delta_{ent}$.} \label{sentconsitency}
\end{figure}

Following \cite{Buchel:2017qwd}, the entanglement entropy
density of a dynamical fixed point $s_{ent}$ is simply the late-time
limit of the non-equilibrium entropy density
\eqref{as}\footnote{See also \cite{Buchel:2019qcq,Buchel:2019pjb}.}:
\begin{equation}
s_{ent}\equiv \lim_{\tau\to \infty} s=\frac{2\pi H^2}{\kappa^2}\
\left(\frac{\calf(x)}{x}\right)^2\bigg|_{x=x_{AH}}\,,
\eqlabel{defsent}
\end{equation}
where we used \eqref{bdfp} for the late-time limit of $\sigma(t,x)$.

Alternatively, the DFP entanglement entropy density can be compute
from \eqref{dasdt}
\begin{equation}
\begin{split}
s_{ent}&=\lim_{\tau\to \infty} \frac{2\pi}{\kappa^2}\ \frac{(\Sigma^2)'}{2H\ e^{2H\tau}}\ \frac{
 (d_+\phi)^2+(d_+\chi)^2}{\phi^2+6-2 \chi^2-g\phi^2 \chi^2}\bigg|_{r=r_{AH}}
\\&=\frac{2\pi H^2}{\kappa^2}\
\biggl[(-x^2\del_x)\left(\frac{\calf(x)}{x}\right)^2\biggr]\ \times\
\frac{\left(-\frac 12 \calg p'\right)^2+\left(-\frac 12 \calg q'\right)^2}
{p^2+6-2q^2-gp^2 q^2}
\bigg|_{x=x_{AH}}\,.
\end{split}
\eqlabel{defsent2}
\end{equation}
An agreement of $s_{ent}^{\eqref{defsent}}$ and  $s_{ent}^{\eqref{defsent2}}$
is a highly nontrivial check on our numerics. In fig.~\ref{sentconsitency}
we plot
\begin{equation}
\delta_{ent}\equiv \frac{s_{ent}^{\eqref{defsent}}}{s_{ent}^{\eqref{defsent2}}}-1\,,
\eqlabel{errsent}
\end{equation}
as a function of $p_1$ for the $\zet_{2}^\chi$-symmetric DFP (the left panel)
and for the symmetry-broken DFP (the right panel). 

Given an analytic solution of the gravitational dual to the $\zet_{2}^\chi$-symmetric DFP in the limit $p_1\to 0$, see \eqref{fpert}-\eqref{gpert}, we find from
\eqref{uah} the location of the apparent horizon as
\begin{equation}
\begin{split}
x_{AH}=&1-\frac16 6^{2/3}\ p_1^{2/3}+\frac{1}{12} 6^{1/3}\ p_1^{4/3}
+\frac19\ p_1^2-\frac{1801}{38880} 6^{2/3}\ p_1^{8/3}+\calo\left(p_1^{10/3}\right)\,.
\end{split}
\eqlabel{xahpert}
\end{equation}
Either from \eqref{defsent} or \eqref{defsent2} we compute
\begin{equation}
\begin{split}
\frac{s_{ent}}{4\pi c H^2}=\frac{1}{2304} 6^{1/3}\ p_1^{4/3}
-\frac{1}{4608}\ p_1^2-\frac{5}{82944} 6^{2/3}\ p_1^{8/3}
+\frac{79}{1866240} 6^{1/3}\ p_1^{10/3}+\calo\left(p_1^4\right) \,.
\end{split}
\eqlabel{sentpertu}
\end{equation}

\section{Numerical setup}\label{num}

We adapt the characteristic formulation of \cite{Chesler:2013lia} for the numerical 
solution of \eqref{evolveoms}-\eqref{coneoms2}.

\subsection{Field redefinitions and the code equations}

We redefine the fields 
\begin{equation}
\{\, \phi\,,\, \chi\,,\, \sigma\,,\, a\,,\, d_+\phi\,,\, d_+\chi\,,\, d_+\sigma\, \}
\ \to\
\{\, p\,,\, q\,,\, s\,,\, \cala\,,\, dp\,,\, dq\,,\, ds\, \}
\eqlabel{fieldsor}
\end{equation}
as follows
\begin{equation}
\begin{split}
&\phi(t,x)=x\ p_1+x\ p(t,x)\;,\\
&\chi(t,x)=x^3\ q(t,x)\;,\\
&\sigma(t,x)=\frac 1x+s(t,x)\;,\\
&a=\frac 12\ \sigma(t,x)^2-\sigma(t,x)+\cala(t,x)\;,\\
&d_+\phi(t,x)=-\frac{p_1}{2}+x\ dp(t,x)\;,\\
&d_+\chi(t,x)=x^3\ dq(t,x)\;,\\
&d_+\sigma(t,x)=x\ ds(t,x)+\frac 12\ \sigma(t,x)^2-\sigma(t,x)+\frac{p_1^2}{16} \;.
\end{split}
\eqlabel{fieldnew}
\end{equation}
Using \eqref{basym}, we find the asymptotic boundary expansion $x\to 0_+$ for the new fields:
\begin{equation}
\begin{split}
&p=p_2(t)\ x + \calo(x^2)\,,\qquad q=q_4(t)\ x+\calo(x^2)\;,\\
&dp=-p_2(t)-p_1 \lambda(t)+p_1+\calo(x)\,,\qquad dq=-2 q_4(t)+\calo(x)\;,\\
&s=\lambda(t)-\frac{p_1^2}{8}\ x+\calo(x^2)\,,\qquad
ds=\mu(t)-\frac{p_1^2}{12}+\frac{p_1}{12}\ p_2(t)+\frac{p_1^2}{12}\ \lambda(t)+\calo(x)\;,\\
&\cala=-\dot\lambda(t)+\left(\mu(t)+\frac{p_1^2}{12}-\frac{p_1}{12}\ p_2(t)-\frac{p_1^2}{12}\ \lambda(t)\right)\
x+\calo(x^2)\;.
\end{split}
\eqlabel{newbasym}
\end{equation}
In new variables \eqref{fieldnew}, the equations of motion used to evolve the system take form:
\begin{equation}
\begin{split}
&\left[\del_{xx}^2+\frac 2x\ \del_x +\frac{x^4}{4}(3q+x
q')^2+\frac{1}{4}(p_1+p+x p')^2\right]s=J_s\;,\\
&J_s\{p,p',q,q'\}=-\frac{x^3}{4}\left(3q+x
q'\right)^2-\frac{1}{4x}(xp'+p+p_1)^2\;,
\end{split}
\eqlabel{ceq1}
\end{equation}
\begin{equation}
\begin{split}
&\left[\del_x+\frac{ x s'+s}{1+x s}\right]\ ds= J_{ds}\;,\\
&J_{ds}\{p,q,s,s'\}=-\frac{s'}{s  x+1}\left(
\frac32 s ^2+\frac{1}{16} p_1^2+\frac{3 s} {x}+\frac{3}{2 x^2}
\right)
-\frac{(s  x+1) (p_1+p )^2}{4x^2}
\\&+\frac{p_1^2}{16x^2 (s  x+1)}
+\frac14 (g x^2 (p_1+p )^2+2) (1+s  x) q ^2 x^2\,,
\end{split}
\eqlabel{ceq2}
\end{equation}
\begin{equation}
\begin{split}
&\left[\del_x+\frac{ x s'+s}{1+x s}\right]\
dp+\left[\frac{x(p_1+p+xp')}{1+x s}\right]\ ds= J_{dp}\;,\\
&J_{dp}\{p,p',q,s,s'\}=-\frac{p'}{16(s  x+1)} \left(8 s ^2 x+p_1^2 x+16 s +\frac8x\right)
+\frac{p_1 s'}{2(s  x+1)}\\
&-\frac{1}{16(s  x+1)} \left(8 s ^2 p +8 s ^2 p_1+p  p_1^2+p_1^3-\frac{8 p} {x^2}\right)
-(p_1+p ) q ^2 x^4 g\,,
\end{split}
\eqlabel{ceq3}
\end{equation}
\begin{equation}
\begin{split}
&\left[\del_x+\frac{2+3 x s+x^2 s'}{x(1+x s)}\right]dq
+\left[\frac{x(q' x +3  q}{1+x s}\right] ds=J_{dq}\;,\\
&J_{dq}\{p,q,q',s\}=-(p_1+p )^2 q  g
-\frac{q'}{16x (1+s  x)} \left(8 s ^2 x^2+p_1^2 x^2+16 s  x+8\right)
\\
&-\frac{q}{16(1+s  x) x^2}  \left(24 s ^2 x^2+3 p_1^2 x^2+80 s  x+56\right)\,,
\end{split}
\eqlabel{ceq4}
\end{equation}
\begin{equation}
\begin{split}
&\left[\del^2_{xx}+\frac 2x\ \del_x\right]\cala
+\biggl[\frac{2(x^2 s'-1)}{x(1+x s)^2}
\biggr] ds-\left[\frac{p_1+p+x p'}{2x}\right]dp
-\left[\frac{x^3(x q'+3 q)}{2}\right]dq=J_\cala\;,\\
&J_\cala\{p,p',q,q',s,s'\}=\frac14 (s  x+1) (s  x-x+1) (p')^2
-\biggl(
\frac{x s (1-s)  (p+p_1)}{2}\\
&-\frac {(p_1+p)(2s-1)}{2}-\frac{2p+p_1}{4x}
\biggr) p'
-(s')^2+\frac{s'}{8(s  x+1)^2} \left(8 s ^2+\frac{8}{x^2}-p_1^2+\frac{16 s}{x}\right)
\\&+\frac{s(p_1+p )^2}{4} \left(s -1+\frac2x\right)
-\frac{(p_1+p ) (x p +p_1 x-p )}{4x^2}
+\frac{p_1^2}{8x^2 (s  x+1)^2}\\
&+\frac14 x^2 (q' x+3 q )^2 (1+s  x) (s  x-x+1)\,,
\end{split}
\eqlabel{ceq5}
\end{equation}
\begin{equation}
\begin{split}
&\dot p=dp -\frac{p_1}{2x}+\frac{p_1+p +x p'}{2} \left(
x (s ^2-2 s +2 \cala )+2 s -2+\frac1x\right)\,,\\
&\dot q=dq +\left(x \left(\frac12 s ^2-s +\cala \right)+s -1+\frac{1}{2 x}\right) (q' x+3 q )\,,\\
&\dot\mu=\frac14 \lambda p_1^2+\frac14 p_1 p_2-3 \mu\,.
\end{split}
\eqlabel{cev}
\end{equation}
Numerical code is organized as follows.
\begin{itemize}
\item {\bf [Step 1]:} assume that at a time step $t$ we have profiles
\begin{equation}
\begin{split}
&\{p(t,x)\,,\, q(t,x)\,,\, p'(t,x)\,,\, q'(t,x) \}\qquad {\rm and}\\
&\{\lambda(t)\,,\, \mu(t)\,,\, p_2(t)\equiv p'(t,x=0)\,,\, q_4(t)\equiv q'(t,x=0)\}
\;.
\end{split}
\eqlabel{nstep1}
\end{equation}
\item  {\bf [Step 2]:} we solve linear in $s$ equation \eqref{ceq1}, subject to boundary conditions
\begin{equation}
s(t,x=0)=\lambda(t)\,,\qquad s'(t,x=0)=-\frac{p_1^2}{8} \;.
\eqlabel{nstep2}
\end{equation}  
\item  {\bf [Step 3]:} we 
solve linear in $ds$ equation \eqref{ceq2}, subject to the boundary conditions
\begin{equation}
ds(t,x=0)=\mu(t)-\frac{p_1^2}{12}+\frac{p_1}{12}\ p_2(t)+\frac{p_1^2}{12}\ \lambda(t) \;.
\eqlabel{nstep3}
\end{equation}  
\item  {\bf [Step 4]:} we 
solve linear in $dp$ equation \eqref{ceq3}, subject to the boundary conditions
\begin{equation}
dp(t,x=0)=-p_2(t)-\lambda(t)\ p_1+p_1\,.
\eqlabel{nstep31}
\end{equation}  
\item  {\bf [Step 5]:} we 
solve linear in $dq$ equation \eqref{ceq4}, subject to the boundary conditions
\begin{equation}
dq(t,x=0)=-2 q_4(t) \;.
\eqlabel{nstep4}
\end{equation}  
\item  {\bf [Step 6]:} we 
solve linear in $a$ equation \eqref{ceq5}, subject to the boundary conditions
\begin{equation}
\cala'(t,x=0)=\mu(t)+\frac{p_1^2(1-\lambda(t))}{12} -\frac{p_1}{12}\ p_2(t)\,,\qquad
\cala(t,x=1)=a^h \;.
\eqlabel{nstep5}
\end{equation}  
The value $a^h$ is determined from the stationarity of the apparent horizon 
at $x=1$ as explained in the following subsection.
\item  {\bf [Step7]:} we use evolution equations \eqref{cev}, along with (see \eqref{newbasym})
\begin{equation}
\dot\lambda(t)=-\cala(t,x=0) \;,
\eqlabel{defdotl}
\end{equation}
to compute 
\begin{equation}
\{p(t+dt,x)\,,\, q(t+dt,x)\,,\,  \lambda(t+dt)\,,\, \mu(t+dt)\} \;.
\eqlabel{nstep6}
\end{equation}
After computing the radial coordinate derivatives $\{p'(t+dt,x)\,,\, q'(t+dt,x)\}$, we 
repeat {\bf [Step 1]}.
\end{itemize}

Notice that the first equation in \eqref{evolveomsx} is redundant in our numerical procedure:
rather than propagating in time $\sigma$, we compute it from the constraint
\eqref{coneomsx} at 
each time step; nonetheless, we monitor the consistency of that equation during the evolution.  

Implementing the code\footnote{Code implementation
is similar to the one used in \cite{Buchel:2014gta}.}, 
we use spectral methods for the radial coordinate integration,
{\bf [Step 2]}- {\bf [Step 6]}. Singularities of the equations at the boundary collocation point 
$x=0$ are resolved using the corresponding boundary conditions instead. We use fourth-order 
Runge-Kutta method for the time evolution, {\bf [Step 7]}.

\subsection{Apparent horizon and the boundary condition for $a$}

Our numerical implementation requires an independent computation of $a^h\equiv
\cala(t,x=1)$ (see \eqref{nstep5}),
given radial profiles $\{p,p',q,q',s,s',dp,ds,dq\}$ and the diffeomorphism
parameter $\lambda$ at time $t$. 
Following \cite{Chesler:2013lia}, this is done by enforcing the time-independent location of the horizon.
Apparent horizon is located as $x=x_{AH}$ such that, see \eqref{ldata}, 
\begin{equation}
\biggl(d_+\sigma(t,x)+\sigma(t,x)\biggr)\bigg|_{x=x_{AH}} =0 \;.
\eqlabel{lochor2}
\end{equation}
Assuming $x_{AH}=1$, $\frac{dx_{AH}}{dt}=0$, and using equations of motion \eqref{ceq1}-\eqref{cev} we compute $a^h$ from
\begin{equation}
\del_t \biggl(d_+\sigma(t,x_{AH})+\sigma(t,x_{AH})\biggr)\bigg|_{x_{AH}=1}=0 \;.
\eqlabel{lochor3} 
\end{equation}
Denoting 
\begin{equation}
\left\{\, p^h\,,\, dp^h\,,\, q^h\,,\, dq^h\,,\, s^h\, \right\}\  \equiv \left\{\, p\,,\, dp\,,\, q\,,\, dq\,,\, s\, \right\}\bigg|_{(t,x=1)}\,,
\eqlabel{defh}
\end{equation}
we find
\begin{equation}
\begin{split}
a^h=&-\frac12 (s^h)^2+\frac12
+\frac{(dq^h)^2 +\left( dp^h-\frac 12p_1\right)^2}{((q^h)^2 g-1) (p^h+p_1)^2+2 (q^h)^2
-6}\,.
\end{split}
\eqlabel{ahres}
\end{equation}

\subsection{Initial conditions}\label{initial}

To evolve \eqref{ceq1}-\eqref{cev} one has to provide data, at $t=0$ 
as required by {\bf [Step 1]}, see \eqref{nstep1}. In particular, 
we need to specify $\lambda_0\equiv \lambda(t=0)$. Once again, we follow \cite{Chesler:2013lia}. 

Recall that both $\phi$ and $\chi$ are left invariant under the reparametrization transformations:
\begin{equation}
\frac 1x\ \to \frac 1x +\lambda_0 \;.
\eqlabel{repx}
\end{equation}
To maintain this invariance, we specify initial conditions 
for $\{p,q\}$ (in $\lambda_0$-invariant way) in terms of 
two amplitudes $\{\cala_p,\cala_q\}$:
\begin{equation}
\begin{split}
&p\bigg|_{t=0}=\cala_p\ \frac{x}{(1+x\lambda_0)^2}\
\exp\left[-\frac{x}{1+x\lambda_0}\right]-\frac{p_1\lambda_0 x}{1+x\lambda_0} \;,\\
&q\bigg|_{t=0}=\cala_q\ \frac{x}{(1+x\lambda_0)^4}\
\exp\left[-\frac{x}{1+x\lambda_0}\right] \;.
\end{split}
\eqlabel{initcond}
\end{equation}
We then proceed as follows\footnote{For this procedure the integration range over the radial coordinate $x$ 
might exceed unity.}:
\begin{itemize}
\item given  $\{\cala_p,\cala_q\}$ we set $\lambda_0=0$ and perform {\bf [Step 2]} \eqref{nstep2}
and    {\bf [Step 3]} \eqref{nstep3};
\item having enough data, we follow \eqref{fieldnew} to compute the profile $d_+\sigma(t=0,x)$;
\item we find numerically the root $x=x_0$ of the equation
\begin{equation}
\biggl(d_+\sigma(t=0,x)+\sigma(t=0,x)\biggr)\bigg|_{x=x_0}=0 \;;
\eqlabel{findx0}
\end{equation}
\item we set the trial value of $\lambda_0$ as
\begin{equation}
\lambda_0=\frac{1}{x_0}-1 \;,
\eqlabel{trial}
\end{equation}
which (apart from the numerical errors) would guarantee that the corresponding location of the 
apparent horizon  is now at $x=1$; 
\item the trial value \eqref{trial} is further adjusted repeatedly performing  {\bf [Step 2]} 
and    {\bf [Step 3]} to achieve  \eqref{findx0}
at a high accuracy.  
\end{itemize}

\subsection{DFP with spontaneously broken $\zet_2$
symmetry as an initial condition}\label{initialb}

Gravitational dual to a DFP with spontaneously broken $\zet_2$
symmetry, see section \ref{z2b}, can be introduced directly
into a numerical code of section \ref{num} as follows:
\begin{itemize}
\item We need to set up {\bf [Step 1]} by providing the profiles  for $p_c(t=0,x)$,
$q_c(t=0,x)$. We used the subscript $ _c$ to differentiate these fields
from the related profiles of a DFP, see \eqref{bdfp}, $p(z)$ and $q(z)$.
Note that we relabeled the radial coordinate used in section \ref{z2b} as 
$x\to z$. The reason for this is that in the evolution code the radial
coordinate varies $x\in [0,1]$ (with the fixed location $x=1$ for the apparent horizon),
while in computing the DFP, the radial coordinate varies as
$z\in [0,\frac 13)\ \cup (\frac 13,x_{AH}]$. We remind the reader that the reason
we are forced to split the integration range in determining the DFP profiles
in section \ref{z2b} is due to the fact that the corresponding equations
\eqref{vb1}-\eqref{vb3} have a coordinate (not physical) singularity whenever $a(t,z)$
vanishes: $a(t,z=x_s)=0$, $x_s\in  (0,x_{AH})$. As done in section \ref{z2b},
it is convenient to keep $x_s$ fixed (our choice is $x_s=\frac 13$) while
to allow for a variation of $x_{AH}$ as one changes $p_1$,
correspondingly $\Lambda$ in \eqref{defformation}. 
\item Radial coordinates used in sections \ref{z2b} and \ref{num} are related by a simple
coordinate diffeomorphism \eqref{resdiffeo}:
\begin{equation}
x=\frac{z}{1+ \delta z}\,,\qquad \delta=1-\frac{1}{x_{AH}}\,.
\eqlabel{xtoz}
\end{equation}
\item The gravitational scalars $\phi$ and $\chi$ are invariant under the
coordinate diffeomorphism \eqref{resdiffeo}; thus, recall \eqref{fieldnew},
\begin{equation}
\begin{split}
&\phi\bigg|_{t=0}=p_1 x + x\ p_c(0,x)\ =\ p\biggl(\frac{x}{1-\delta x}\biggr)\,,\qquad x\in [0,1]\,, \\ 
&\chi\bigg|_{t=0}=x^3\ q_c(0,x)\ =\ q\biggl(\frac{x}{1-\delta x}\biggr)\,,\qquad x\in [0,1]  \,.
\end{split}
\eqlabel{relationphichi}
\end{equation}
Given Wolfram Mathematica computed profiles $p(z)$ and $q(z)$, it is straightforward to use
\eqref{xtoz} and \eqref{relationphichi} to output the profiles $p_c(0,x)$ and $q_c(0,x)$
at collocation points, suitable for FORTRAN code used in section \ref{num}. 
\item The diffeomorphism parameters $\lambda_c(t=0)$ (used in section \ref{num})
and the corresponding parameter $\lambda$ (used in section \ref{z2b}) are related as
\begin{equation}
\lambda_c(t=0)=\lambda-\delta\,.
\eqlabel{lambdarel}
\end{equation}
\item From \eqref{relationphichi},
\begin{equation}
p_{2,c}(t=0)=p_2 +p_1 \delta\,.
\eqlabel{p2rel}
\end{equation}
where again we used the subscript $ _c$ to differentiate the corresponding
parameters used in sections \ref{num} and \ref{z2b}.
\item To complete set up of {\bf [Step 1]} we need $\mu_c(t=0)$. Since the input is
a DFP,
\begin{equation}
\dot\mu_c(t=0)=0\,,
\eqlabel{dmu}
\end{equation}
thus, from \eqref{cev}
\begin{equation}
\mu_c(t=0)=\frac {1}{12} \lambda_c\ p_1^2 + \frac{1}{12}\ p_1\ p_{2,c}=
\frac{1}{12}\ p_1 \left(p_{2,c}+\lambda_c p_1\right)= \frac{1}{12}\ p_1 \left(p_{2}+\lambda p_1\right)\,,
\eqlabel{mucinput}
\end{equation}
where to arrive at  the last equality we used \eqref{p2rel}.
\end{itemize}

The manipulations described above necessarily introduce numerical noise --- which is important
given that $\zet_2$ spontaneously broken DFPs are perturbatively unstable.

\subsection{Convergence tests}\label{convtests}

\begin{figure}[h]
\begin{center}
\psfrag{a}[cc][][0.6][0]{$\qquad\qquad N=60$}
\psfrag{b}[cc][][0.6][0]{$\qquad\qquad N=70$}
\psfrag{c}[cc][][0.6][0]{$\qquad\qquad N=80$}
\psfrag{t}[cc][][1][0]{$t=\tau H$}
\psfrag{q}[bb][][1][0]{$\frac{\calo_\chi}{cH^4}$}
\psfrag{k}[tt][][1][0]{$\ln\calk_{AH}$}
\includegraphics[width=3in]{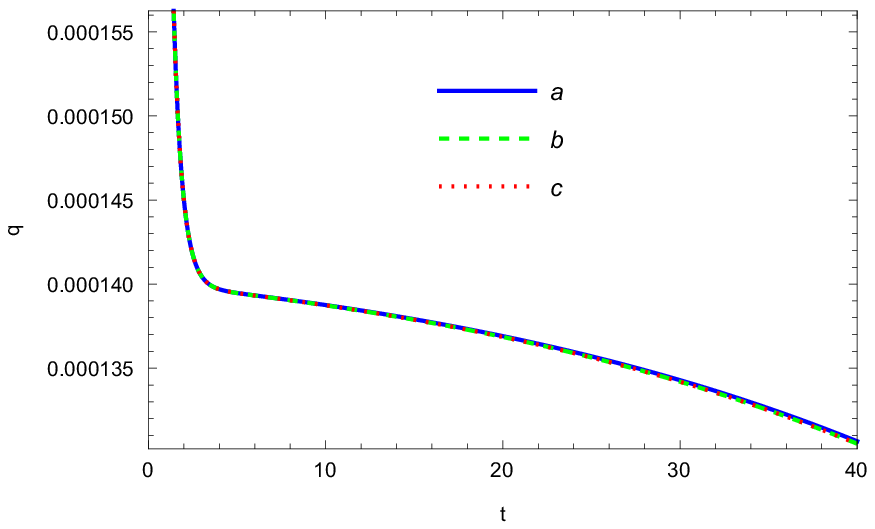}
\includegraphics[width=3in]{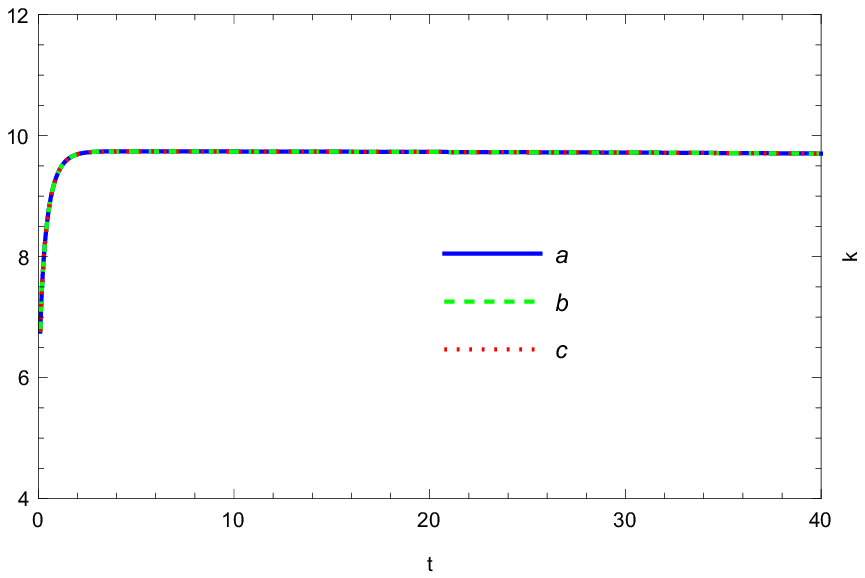}
\end{center}
  \caption{Convergence of simulations performed at different collocation points $N$
  with a DFP attractor at late times. Here, $\frac{\Lambda}{H}=p_1^*$, see \eqref{pqdef}.
} \label{conv05566}
\end{figure}

\begin{figure}[h]
\begin{center}
\psfrag{a}[cc][][0.6][0]{$\qquad\qquad N=60$}
\psfrag{b}[cc][][0.6][0]{$\qquad\qquad N=70$}
\psfrag{c}[cc][][0.6][0]{$\qquad\qquad N=80$}
\psfrag{t}[cc][][1][0]{$t=\tau H$}
\psfrag{q}[bb][][1][0]{$\ln \frac{\calo_\chi}{cH^4}$}
\psfrag{k}[tt][][1][0]{$\ln\calk_{AH}$}
\includegraphics[width=3in]{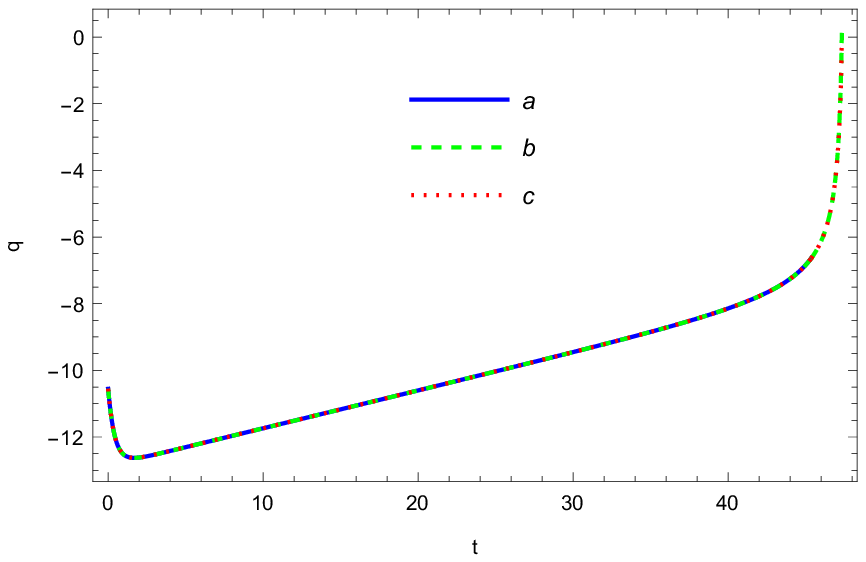}
\includegraphics[width=3in]{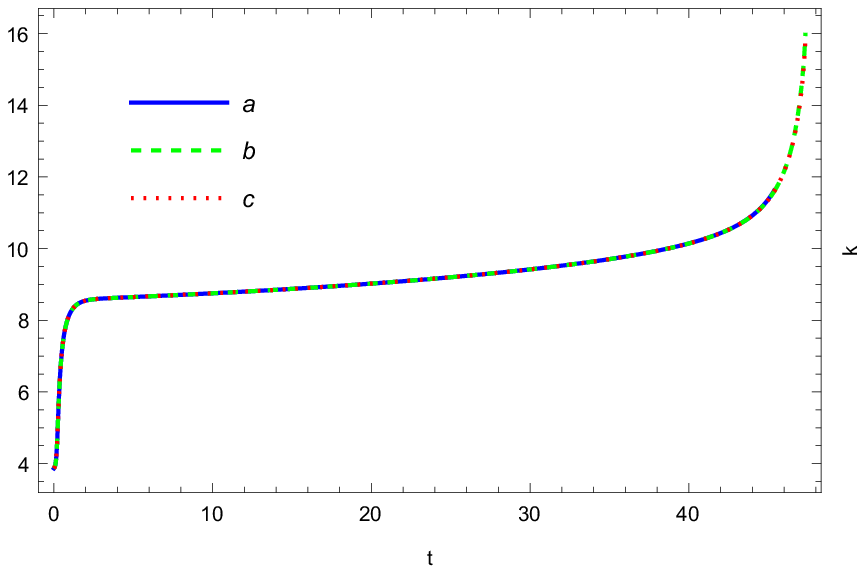}
\end{center}
  \caption{Convergence of simulations performed at different collocation points $N$
  without an  attractor at late times. Here, $\frac{\Lambda}{H}=\tp_1$, see \eqref{pqdef2}.
} \label{div057}
\end{figure}

We performed self-convergence tests to verify the validity of the
obtained numerical solutions. In particular, we study each
configuration numerically under different number of collocation points
$N=60, 70, 80$.  We monitored the convergence of the residuals of the
constraint equations to zero as well as each evolved field (and
computing self-convergence test by a suitable interpolation onto a
finite difference grid). Additionally, we confirmed convergence
of the Kretschmann scalar $\calk$ at the
apparent  horizons, $\calk_{AH}=\calk(t,x_{AH})$:
\begin{equation}
\begin{split}
\calk(t,x)\equiv R_{\mu\nu\rho\lambda} R^{\mu\nu\rho\lambda} =&
x^4 \biggl(
\chi'\ d_+\phi+\phi'\ d_+\chi
\biggr)^2
+2 x^4 \biggl(
(\chi'\ d_+\chi)^2+(\phi'\ d_+\phi)^2\biggr)\\
&+(\phi^2 \chi^2 g-\phi^2+2 \chi^2-6)^2\,,
\end{split}
\eqlabel{defk}
\end{equation}
where we used the equations of motion \eqref{evolveomsx} and \eqref{coneomsx}.
As an illustration,
figs.~\ref{conv05566} and \ref{div057} display the expectation
values $\calo_\chi(t)$ and  $\calk_{\rm AH}(t)$ for both
the stable and unstable configurations correspondingly.
The fractional difference between the time series of the observables \eqref{vev1}-\eqref{vev4},
\eg
\begin{equation}
\max_t\  \bigg|\frac{\calo^{N_2}(t)}{\calo^{N_1}(t)}-1\bigg|\,,
\eqlabel{fracdif}
\end{equation}
for the runs with $N_1=70$ and $N_2=80$ collocation points is $\sim 10^{-6}-10^{-5}$,
both for stable and unstable configurations. All the numerical runs reported in the
body of the paper are performed with $N=80$ collocation points.


\bibliographystyle{JHEP}
\bibliography{dfp}

\end{document}